\documentclass[12pt]{iopart}

\usepackage{iopams}
\usepackage{graphicx}
\usepackage{bm}
\usepackage{cite}
\usepackage{color}

\graphicspath{{.}{./EPS/}{./Figures/}}

\newcommand{\ket}[1]{\left | #1 \right \rangle}
\newcommand{\bra}[1]{\left \langle #1 \right |}

\newcommand*{\ee}{\ensuremath{\mathrm{e}}}

\newcommand*{\Bb}{\ensuremath{\mathcal{B}}}
\newcommand*{\vek}[1]{{\ensuremath{\bm{\mathrm{#1}}}}}

\newcommand*{\rr}{{\ensuremath{\bm{\mathrm{r}}}}}

\begin{document}

\title[Fragility of the fractional quantum spin Hall effect]{Fragility of the fractional quantum
spin Hall effect in quantum gases}

\author{O. Fialko$^1$, J.  Brand$^1$, U. Z\"ulicke$^2$}

\address{$^1$ Institute of Natural and Mathematical Sciences and Centre for
Theoretical Chemistry and Physics, Massey University, Auckland 0632, New
Zealand}

\address{$^2$ School of Chemical and Physical Sciences and MacDiarmid Institute
for Advanced Materials and Nanotechnology, Victoria University of Wellington,
PO Box 600, Wellington 6140, New Zealand}

\ead{uli.zuelicke@vuw.ac.nz}

\begin{abstract}
We consider the effect of contact interaction in a prototypical quantum spin Hall system of pseudo-spin-1/2 particles. A strong effective magnetic field with opposite directions for the two spin states restricts two-dimensional particle motion to the lowest Landau level.
While interaction between same-spin particles leads to incompressible correlated states at 
fractional filling factors as known from the fractional quantum Hall effect, these states are 
destabilized by interactions between opposite spin particles. Exact results for two particles with 
opposite spin reveal a quasi-continuous spectrum of extended states with a large density of 
states at low energy. This has implications for the prospects of realizing the fractional quantum 
spin Hall effect in electronic or ultra-cold atom systems. Numerical diagonalization is used to
extend the two-particle results to many bosonic particles and trapped systems. The interplay
between an external trapping potential and spin-dependent interactions is shown to open up
new possibilities for engineering exotic correlated many-particle states with ultra-cold atoms.
\end{abstract}


\submitto{\NJP}

\maketitle

\section{Introduction}

Trapped ultra-cold atoms have become model systems of choice for simulating physical
effects from condensed matter~\cite{blo12} to cosmology~\cite{hun13,opa13}. The recently
achieved ability to create synthetic vector potentials~\cite{dal11} acting on neutral atoms
has increased the versatility of the atomic-physics simulation toolkit even further. It is now
possible to simulate magnetic fields by inducing spatially varying $U(1)$ (i.e., scalar)
gauge potentials~\cite{lin09a,lin09b,jim12}, and pseudo-spin splittings can be created in
spinor gases~\cite{lin11,gal13} using spatially constant vector potentials having a (possibly
non-Abelian) matrix structure. These advances have stimulated a host of theoretical works
studying, e.g., the effect of uniform $SU(2)$ gauge potentials on the behavior of quantum
particles subject to uniform ordinary magnetic fields~\cite{gol09a,bur10,est11,pal11}, or
proposing the use of staggered effective spin-dependent magnetic fields in optical
lattices~\cite{gol10,ber11,mei12,maz12} to simulate a new class of materials called
topological insulators~\cite{has10,has11,qi11} that exhibit the quantum spin Hall
effect~\cite{kan05,ber06a,ber06b,kon07}. Very recently, the non-quantized intrinsic spin
Hall effect~\cite{mur03,sin04,zhu06,liu07} has been realized experimentally in a quantum
gas~\cite{bee13}, and the authors of this paper outline the way forward to reaching
conditions where the quantum spin Hall effect could be observed. Furthermore, newly
demonstrated methods to simulate strong-enough magnetic fields to probe ultra-cold
atom gases in the ordinary quantum-Hall regime~\cite{aid13,miy13} are expected to be
adaptable for the purpose of generating spin-dependent quantizing magnetic
fields~\cite{aid13,ken13}, which opens up another avenue towards the exploration of
quantum-spin-Hall physics. Part of the motivation for our present theoretical work arises
from these rapid developments of experimental capabilities.

The ordinary quantum Hall (QH) effect~\cite{pra90} occurs because particles confined to
move in two spatial dimensions and subject to a strong perpendicular magnetic field
develop incompressibilities at integer, and certain fractional, values of the Landau-level
filling factor~\cite{mac95}. While the Landau quantization of single-particle energies is the
origin of the integer QH effect, incompressibility at fractional filling factors is caused by the
discrete spectrum of interaction energies for two particles occupying states from the same
Landau level~\cite{hal90,coo08,vie08}. In the conceptually simplest realization of the
quantum \emph{spin\/} Hall (QSH) effect~\cite{ber06a}, particles exhibit an integer QH
effect due to a spin-dependent perpendicular magnetic field that points in opposite
directions for the two opposite-spin components. It seems then quite straightforward to
conjecture~\cite{ber06a,liu09} that a fractional version of the QSH effect should exist that
mirrors features of the ordinary fractional QH effect in multi-component
systems~\cite{hal83a,rei02,rei04}. Following this line of thought, some previous discussions
of a putative fractional QSH physics~\cite{liu09,lan12} have been based on an \textit{ad hoc\/}
adaptation of trial wave functions first proposed in Ref.~\cite{ber06a}. However, unless only
particles with the same spin interact, such an approach is fraught with difficulty~\cite{goe12}.

Here we revisit the question of how a fractional QSH effect can arise in an interacting
(pseudo-)spin-1/2 system that experiences a spin-dependent quantizing magnetic
field. In particular, we elucidate the effect of interactions between particles having
opposite spin. We find that such inter-species interactions significantly alter the
expected QSH physics, but they also open up new opportunities for tailoring the properties
of quantum many-particle states. Our study is complementary to recent investigations of
fractional QSH phases~\cite{lev09,lev11,qi11a,neu11a,goe12} that arise in materials with
exotic topological band structures~\cite{hal88,tan11,sun11,neu11} or strained
graphene~\cite{gha12}. The results obtained here are relevant for electronic systems as
well as for ultra-cold bosonic or fermionic atoms. Particular attention is paid to trapped bosons.

The article is organized as follows. In Sec.~\ref{sec:Hamiltonian}, we introduce the
basic model description of an interacting system of (pseudo-)spin-1/2 particles that are
subject to a spin-dependent magnetic field. The single-particle states are given in the
representation of spin-dependent guiding-center and Landau-level quantum numbers.
The eigenvalue problem of two interacting particles is solved -- for both cases of equal
and opposite-spin particles -- in the subsequent Sec.~\ref{sec:TwoPart}. When the two
interacting particles have opposite spin, important differences arise with respect to the
classic results obtained~\cite{hal83} for spinless (or same-spin) particles. We explore the
ramifications of this fact by numerical exact-diagonalization studies with up to 6 bosons
for which results are presented in Sec.~\ref{sec:numFew}. Our conclusions are summarized
in Sec.~\ref{sec:conc}.

\section{Pseudo-spin-1/2 particles subject to a spin-dependent magnetic field
\label{sec:Hamiltonian}}

We consider a gas of particles (e.g., atoms) that carry a (pseudo-)spin-1/2 degree
of freedom and are confined to move in the $xy$ plane. A spin-dependent vector
potential
\begin{equation}\label{eq:ACgauge}
\vek{\mathcal A}(\vek{r}) = \frac{\Bb}{2}\left( -y\, \vek{\hat x} + x\, \vek{\hat{y}} \right)
\sigma_z \equiv \left( \begin{array}{cc} \vek{\mathcal A}^{(+)}(\vek{r}) & 0 \\ 0 &
\vek{\mathcal A}^{(-)}(\vek{r}) \end{array} \right) 
\end{equation}
is presumed to be generated (e.g., by optical means in an atom
gas~\cite{dal11,bee13,aid13,ken13}). Here $\sigma_z$ denotes the diagonal Pauli
matrix, and the $\vek{\hat \jmath}$ are Cartesian unit vectors in real space.

The vector potential (\ref{eq:ACgauge}) is Abelian and gives rise to a spin-dependent
magnetic field perpendicular to the $xy$ plane: $\vek{\mathcal B} \equiv\vek{\nabla}
\times\vek{\mathcal A} = \Bb\, \vek{\hat z}\, \sigma_z$. This situation of opposite-spin
particles being subjected to oppositely directed magnetic fields corresponds directly to
setups considered for a semiconductor heterostructure~\cite{ber06a,xu11} and in
neutral-atom systems~\cite{zhu06,liu07,bee13,ken13}. A somewhat related study in the
context of cold bosonic gases was given in Ref.~\cite{jul12}, only that there the two spin
components also experience a large Zeeman-like energy shift and, therefore, this work
focused only on the dynamics of a single component.

Without loss of generality, we will assume $\Bb>0$ from now on. Also note that, with
unit conventions chosen in this article, the `magnetic-field' magnitude $\mathcal B$
is related to a fundamental (`magnetic') length scale $l_{\mathcal B} = \sqrt{\hbar/
{\mathcal B}}$.

\subsection{Many-particle model Hamiltonian}

The most general form of the many-body Hamiltonian that describes our system of
interest is $\mathcal H = {\mathcal H}_0 + {\mathcal H}_{\mathrm{int}}$, where
\numparts
\begin{eqnarray}
{\mathcal H}_0 &=& \int d^2 r \,\, \hat\Psi^\dagger(\rr) \left\{ \frac{1}{2 M} \left[
\vek{p} \, \hat 1 - \vek{\mathcal A}(\vek{r}) \right]^2 + {\mathcal V}(\vek{r})
-\mu \, \hat 1 \right\} \hat\Psi(\rr) \quad , \\ \label{eq:model_int}
{\mathcal H}_{\mathrm{int}} &=& \int d^2 r \,\, \sum_{\sigma, \sigma^\prime = \pm}\,
\frac{g_{\sigma\sigma^\prime}}{2} \, \hat\Psi_{\sigma}^\dagger(\rr)
\hat\Psi_{\sigma^\prime}^\dagger(\rr) \hat\Psi_{\sigma^\prime}(\rr)\hat\Psi_{\sigma}(\rr)
\quad .
\end{eqnarray}
\endnumparts
Here $\hat 1$ denotes the $2\times 2$ unit matrix, and $\hat\Psi(\rr)\equiv \Big(\hat
\Psi_+(\rr),\hat\Psi_-(\rr) \Big)^T$ is a two-spinor of (bosonic or fermionic) annihilation
operators for a particle located at position $\vek{r}$ and having a definite pseudo-spin-1/2
projection along the $\vek{\hat z}$ axis. Furthermore, the relation $g_{+-}=g_{-+}$ is
implicit in the formalism. As it is often useful, we also give an equivalent expression for
the interaction part of the Hamiltonian,
\numparts
\begin{eqnarray}
{\mathcal H}_{\mathrm{int}} &=& {\mathcal H}_{\mathrm{int}}^{(0)} +
{\mathcal H}_{\mathrm{int}}^{(1)} + {\mathcal H}_{\mathrm{int}}^{(2)} \quad , \\
\label{eq:ints_1}
{\mathcal H}_{\mathrm{int}}^{(0)} &=&  \frac{g_0}{2} \int d^2 r \,\,\, \sum_\sigma
\hat\Psi^\dagger_\sigma \left(\hat\Psi^\dagger \hat\Psi\right) \hat \Psi_\sigma \quad , \\
{\mathcal H}_{\mathrm{int}}^{(1)} &=& \frac{g_1}{2} \int d^2 r \,\,\, \sum_\sigma \sigma\,
\hat\Psi_{\sigma}^\dagger\hat\Psi_{\sigma}^\dagger \hat\Psi_{\sigma} \hat\Psi_{\sigma}
\quad , \\ {\mathcal H}_{\mathrm{int}}^{(2)} &=& \frac{g_2}{2} \int d^2 r \,\,\, \sum_\sigma
\sigma \hat\Psi^\dagger_\sigma \left(\hat\Psi^\dagger \sigma_z \hat\Psi \right) \hat
\Psi_\sigma \quad ,
\label{eq:ints_2}
\end{eqnarray}
\endnumparts
with $g_0 = \sum_{\sigma\sigma^\prime} g_{\sigma\sigma^\prime}/4$, $g_1 = (g_{++} -
g_{--})/2$, and $g_2 = (g_{++} + g_{--}-2\, g_{+-})/4$. Depending on whether the asymmetric
interaction couplings $g_1$, $g_2$ are positive or negative, different physical regimes may be
reached~\cite{ho11,ram12}.\footnote{The notation used in Eqs.~(\ref{eq:ints_1})--(\ref{eq:ints_2}) 
can be related to that which is often adopted in the atom-gas literature~\cite{ho08,wan10} by
setting $g_0\equiv c_0$, $g_2\equiv c_2$, and $g_1\equiv 0$. See also Ref.~\cite{ohm98}.
Note, however, the different parameterization used in Ref.~\cite{lin11} where $c_{0,2}$ are
interaction constants associated with the atomic spin-1 degree of freedom from which the
pseudo-spin-1/2 components are derived. Our notation is related to theirs via $g_0\equiv c_0
+\frac{3}{4} c_2 + \frac{1}{4} c^\prime_{\uparrow\downarrow}$, $g_1 \equiv -\frac{1}{2} c_2$,
and $g_2\equiv -\frac{1}{4} (c_2 + c^\prime_{\uparrow\downarrow})$.}

In the case where $g_{+ -} = 0$, the system reduces to two independent two-dimensional
(electron or atom) gases that are each subject to a perpendicular magnetic field. Known
phenomena associated with the fractional QH effect~\cite{pra90,mac95,coo08,vie08}
will then be exhibited by the individual systems. However, as seen from our study presented
in Secs.~\ref{sec:TwoPart} and \ref{sec:numFew} below, the behavior of the system with
$g_{+-}\ne 0$ departs from the previously considered~\cite{hal83a} two-component
fractional-QH physics because of the very different type of constraints that is placed on
the orbital motion of particles subject to oppositely directed magnetic fields. 

\subsection{Spin-dependent single-particle Landau levels}

We now consider single-particle states associated with spin component
$\sigma$. The kinetic momentum is $\vek{\pi}^{(\sigma)} = \vek{p}-
\vek{\mathcal A}^{(\sigma)}$, and straightforward calculation establishes the
commutation relations
\begin{equation}
[\pi_\alpha^{(\sigma)},\pi_{\alpha'}^{(\sigma')}]=\sigma\, i \,\frac{\hbar^2}{l_{\mathcal B}^2}
\, \delta_{\sigma\sigma'}\, \epsilon_{\alpha\alpha'} \quad ,
\end{equation}
where $\epsilon_{xy}=-\epsilon_{yx}=1$. Operators for the guiding-center locations can
then be defined in the usual manner~\cite{mac95}, $\vek{R}^{(\sigma)}= \vek{r}- \sigma\,
l_{\mathcal B}^2\, [\vek{\hat{z}}\times\vek{\pi}^{(\sigma)}]/\hbar$, and their components
satisfy the commutation relations
\begin{equation}
[R^{(\sigma)}_\alpha, R^{(\sigma')}_{\alpha'}]=-\sigma\, i \, l_{\mathcal B}^2 \,
\delta_{\sigma\sigma'}\,\epsilon_{\alpha\alpha'} \quad .
\end{equation} 
Moreover, we find $[R^{(\sigma)}_\alpha, \pi_{\alpha'}^{(\sigma')}]=0$. Following the
familiar approach~\cite{mac95}, we define the harmonic-oscillator Landau-level ladder
operator for states with spin $\sigma$ via
\begin{equation}
a_\sigma = \frac{i\, l_{\mathcal B}}{\sqrt{2}\,\hbar}\, \left( \pi^{(\sigma)}_x + \sigma\, i\,
\pi^{(\sigma)}_y \right) \quad .
\end{equation}
Similarly, the ladder operator operating within a Landau level for spin component
$\sigma$ is
\begin{equation}
b_\sigma=\frac{1}{\sqrt{2}\, l_{\mathcal B}}\, \left( R^{(\sigma)}_x - \sigma\, i \,
R^{(\sigma)}_y \right) \quad .
\end{equation}

We can express the kinetic energy and the $z$ component of angular momentum in
terms of the ladder operators [$\omega_{\mathrm{c}} = \hbar/(M l_{\mathcal B}^2)$]:
\numparts
\begin{eqnarray}
\frac{\left[\vek{\pi}^{(\sigma)}\right]^2}{2 M} &=& \hbar\omega_{\mathrm{c}} \left(
a_\sigma^\dagger a_\sigma + \frac{1}{2} \right) \quad , \\ \label{eq:angMom}
L_z &\equiv& x\, p_y - y\, p_x = \hbar\, \sigma \left( b_\sigma^\dagger b_\sigma -
a_\sigma^\dagger a_\sigma \right) \quad .
\end{eqnarray}
\endnumparts
Landau-level eigenstates are generated via
\begin{equation}
\ket{\{n_\sigma, m_\sigma\}} = \left[ \sum_\sigma \frac{\left( a_\sigma^\dagger
\right)^{n_\sigma}}{\sqrt{n_\sigma!}} \, \frac{\left( b_\sigma^\dagger\right)^{m_\sigma}}
{\sqrt{m_\sigma !}} \, {\mathcal P}_\sigma\right] \ket{\mathrm{vac}} \,\, ,
\end{equation}
where $\ket{\mathrm{vac}} = (1, 1)^T \ket{0}$ and $\ket{0}$ is the state that is annihilated by
all ladder operators $a_\sigma$ and $b_\sigma$. In the following, we will focus on
the case where all particles are in the lowest Landau level, i.e., when $n_+=n_-=0$.

\section{Two-particle interactions in the lowest Landau level: Spin matters
\label{sec:TwoPart}}

\subsection{Heuristic real-space picture}

Before presenting a formal analysis of the interacting two-particle system subject to a
strong spin-dependent magnetic field in the following Subsection, we provide a heuristic
argument for how the cases where the two particles feel the same and opposite
magnetic fields differ.

Consider two particles, located at $\vek{r}_1$ and $\vek{r}_2$, respectively, that interact
via a generic potential $V(\vek{r}_1 - \vek{r}_2)$. The corresponding first-quantized
two-particle Hamiltonian reads
\begin{equation}
{\mathcal H}_{1, 2} = \frac{\left[ \vek{p}_1 - \vek{\cal{A}}^{(\sigma_1)}(\vek{r}_{1})
\right]^2}{2M}  + \frac{\left[ \vek{p}_2 - \vek{\cal{A}}^{(\sigma_2)}(\vek{r}_{2})\right]^2}{2M}
+ V(\vek{r}_1 - \vek{r}_2) \quad ,
\end{equation}
with the spin-dependent vector potentials from Eq.~(\ref{eq:ACgauge}). It is straightforward
to show that the sum of kinetic-energy contributions for each particle can be re-arranged in
terms of the linear combinations
\begin{equation}
\vek{R}_{\sigma_1 \sigma_2} = \frac{1}{2} \left( \begin{array}{c} x_1 + x_2 \\ y_1 + \sigma_1
\sigma_2\, y_2 \end{array} \right)\,\, , \,\, \vek{r}_{\sigma_1\sigma_2} = \frac{1}{2} \left(
\begin{array}{c} x_1 - x_2 \\ y_1 - \sigma_1 \sigma_2\, y_2 \end{array} \right)\,\, ,
\end{equation}
yielding the expression
\begin{eqnarray}
{\mathcal H}_{1, 2} &=& \frac{\left[
\vek{P}_{\sigma_1 \sigma_2} - 2 \vek{\cal{A}}^{(+)}(\vek{R}_{\sigma_1\sigma_2})
\right]^2}{2M} + \frac{\left[\vek{p}_{\sigma_1 \sigma_2} - 2 \vek{\cal{A}}^{(+)}
(\vek{r}_{\sigma_1\sigma_2})\right]^2}{2 M}+V(\vek{r}_1 - \vek{r}_2) . \nonumber \\
\end{eqnarray}

The two-particle problem for particles with the same spin reduces to two independent
single-particle problems in the center-of-mass (COM) and relative-coordinate degrees of
freedom ($\vek{R}_{\sigma\sigma}$ and $\rr_{\sigma\sigma}$, respectively) because
$\rr_1 - \rr_2 \equiv 2 \rr_{\sigma \sigma}$. In that case, only the relative-coordinate
degree of freedom feels the interaction potential $V(\rr_{\sigma\sigma})$, and it can be
minimized by placing two particles away from each other. In the specific case of $V(\vek{r})
\propto\delta(\vek{r})$, the zero-energy states of the two-particle system are of the form
$\psi_{\sigma\sigma}(\vek{r}_1,\vek{r}_2)\propto (z_1+z_2)^{m_\mathrm{C}}(z_1-
z_2)^{m_\mathrm{r}}$, where $z_j = x_j + i\, y_j$ is a commonly used complex notation
for the position of particle $j$~\cite{mac95}. The non-negative integers $m_\mathrm{C}$ and
$m_\mathrm{r}$ correspond to the quantized values of COM angular momentum and relative
angular momentum, respectively~\cite{mac95}.

Quite a different situation arises for opposite-spin particles. The kinetic energy of the
two-particle system decouples in the coordinates $\vek{R}_{+-}$ and $\rr_{+-}$,
motivating the proposal of trial wave functions~\cite{ber06a} $\psi_{+-}(\vek{r}_1,
\vek{r}_2) \propto (z_1+z_2^\ast)^{m_\mathrm{C}}(z_1-z_2^\ast)^{m_\mathrm{r}}$.
However, $V(\vek{r})$ still couples the two-particle coordinates $\vek{R}_{+-}$ and
$\rr_{+-}$ and, as a result, the proposed wave function is energetically not favorable for
interacting particles~\cite{goe12}. Rigorous examination of the interacting two-particle
system in the opposite-spin configuration (see below) shows that energy eigenstates
are not eigenstates of COM angular momentum or relative angular momentum and,
furthermore, have an unusual distribution.

\subsection{Rigorous analysis of interaction matrix}

To gain a deeper understanding of the effect of two-particle interactions, we follow the
basic approach employed by previous studies of the fractional QH
effect~\cite{hal90,mac95} and find the interaction potential in the representation of
lowest-Landau-level states. The starting point of such an analysis is the Fourier
decomposition of a spin-dependent interaction potential given by
\begin{equation}\label{eq:Fourierpot}
V_{\sigma_1\sigma_2}(\vek{r}_1 - \vek{r}_2) = \int \frac{d^2 q}{(2\pi)^2} \,\, V_{\sigma_1
\sigma_2}(\vek{q}) \,\, \ee^{i \vek{q}\cdot (\vek{r}_1 - \vek{r}_2)} \quad ,
\end{equation}
because its matrix elements can then be directly related to the corresponding matrix
elements of the exponential in the integrand of (\ref{eq:Fourierpot}). To find the latter for
states in the lowest Landau level, we use the relation
\begin{equation}
\vek{r}_1 - \vek{r}_2 = \vek{R}_{1}^{(\sigma_1)} - \vek{R}_{2}^{( \sigma_2)} +
\frac{l_{\mathcal B}^2}{\hbar} \, \vek{\hat z}\times \left(\sigma_1\,
\vek{\pi}_{1}^{(\sigma_1)} - \sigma_2\, \vek{\pi}_{2}^{(\sigma_2)}\right)
\end{equation}
and the straightforwardly obtained expressions
\numparts
\begin{eqnarray}
&& \vek{q}\cdot \left( \vek{R}_{1}^{(\sigma_1)} - \vek{R}_{2}^{(\sigma_2)} \right) =
\frac{l_{\mathcal B}}{\sqrt{2}} \left[ q_x \left( b_{\sigma_1} - b_{\sigma_2} +
b_{\sigma_1}^\dagger - b_{\sigma_2}^\dagger\right)
\right. \nonumber \\ && \hspace{4.5cm} \left.
+ \sigma_1 i \, q_y \left( b_{\sigma_1} - \sigma_1\sigma_2\, b_{\sigma_2} - 
b_{\sigma_1}^\dagger + \sigma_1\sigma_2 \, b_{\sigma_2}^\dagger\right) \right] \, , \\ 
&& \frac{l_{\mathcal B}^2}{\hbar}\left(\vek{q} \times \hat{\vek{z}}\right)\cdot \left(
\sigma_1\, \vek{\pi}_{1}^{(\sigma_1)} - \sigma_2\, \vek{\pi}_{2}^{(\sigma_2)}\right) =
\frac{l_{\mathcal B}}{\sqrt{2}} \left[ q_x \left( a_{\sigma_1} - a_{\sigma_2} +
a_{\sigma_1}^\dagger - a_{\sigma_2}^\dagger\right)
\right. \nonumber \\ && \hspace{4.5cm} \left.
+ \sigma_1 i \, q_y \left( a_{\sigma_1}
 - \sigma_1\sigma_2\, a_{\sigma_2} - a_{\sigma_1}^\dagger +\sigma_1 \sigma_2 \,
 a_{\sigma_2}^\dagger\right) \right] . 
\end{eqnarray}
\endnumparts
Introducing spin-resolved ladder operators for COM and relative angular momentum,
\begin{equation}
b^{\mathrm{(C)}}_{\sigma_1\sigma_2} = \frac{b_{1\sigma_1} + b_{2\sigma_2}}{\sqrt{2}}\, , \ \
b^{\mathrm{(r)}}_{\sigma_1\sigma_2} = \frac{b_{1\sigma_1} - b_{2\sigma_2}}{\sqrt{2}}\, ,
\end{equation}
and analogous ladder operators for COM energy and relative-motion energy,
\begin{equation}
a^{\mathrm{(C)}}_{\sigma_1\sigma_2} = \frac{a_{1\sigma_1} + a_{2\sigma_2}}{\sqrt{2}}\, , \ \
a^{\mathrm{(r)}}_{\sigma_1\sigma_2} = \frac{a_{1\sigma_1} - a_{2\sigma_2}}{\sqrt{2}}\, ,
\end{equation}
we find using $q_\sigma = q_x + \sigma i\, q_y$
\numparts
\begin{eqnarray}\label{eq:spinExp1}
\ee^{i\vek{q}\cdot \left(\vek{R}_{1}^{(\sigma_1)} - \vek{R}_{2}^{(\sigma_2)}\right)} =
\ee^{-\frac{q^2\, l_{\mathcal B}^2}{2}} \nonumber \\[0.2cm] \times\left\{ \begin{array}{cl} \ee^{i
\bar q_\sigma l_{\mathcal B}\, b^{\mathrm{(r)}\dagger}_{\sigma\sigma}}\,\, \ee^{i
q_\sigma l_{\mathcal B}\, b^{\mathrm{(r)}}_{\sigma\sigma}} & \mbox{for}\, \sigma_1
=\sigma_2 \equiv \sigma \\ \ee^{i q_x l_{\mathcal B}\, b^{\mathrm{(r)}\dagger}_{\sigma,
-\sigma}}\,\, \ee^{i q_x l_{\mathcal B} \, b^{\mathrm{(r)}}_{\sigma,-\sigma}}\,\, \ee^{\sigma
q_y l_{\mathcal B}\, b^{\mathrm{(C)}\dagger}_{\sigma,-\sigma}}\,\, \ee^{-\sigma q_y
l_{\mathcal B}\, b^{\mathrm{(C)}}_{\sigma,-\sigma}}  & \mbox{for} \, \sigma_1= -\sigma_2
\equiv \sigma \end{array} \right. , \\[0.1cm]
\ee^{i\frac{l_{\mathcal B}^2}{\hbar}\left(\vek{q} \times \hat{\vek{z}}\right)\cdot \left(
\sigma_1\, \vek{\pi}_{1}^{(\sigma_1)} - \sigma_2\, \vek{\pi}_{2}^{(\sigma_2)}\right)} =
\ee^{-\frac{q^2\, l_{\mathcal B}^2}{2}}  \nonumber\\[0.2cm] \times \left\{ \begin{array}{cl}
\ee^{i \bar q_\sigma
l_{\mathcal B}\, a^{\mathrm{(r)}\dagger}_{\sigma\sigma}}\,\, \ee^{i q_\sigma l_{\mathcal B}\,
a^{\mathrm{(r)}}_{\sigma\sigma}} & \mbox{for}\, \sigma_1=\sigma_2 \equiv \sigma \\
\ee^{i q_x l_{\mathcal B}\, a^{\mathrm{(r)}\dagger}_{\sigma,-\sigma}}\,\, \ee^{i q_x
l_{\mathcal B} \, a^{\mathrm{(r)}}_{\sigma,-\sigma}}\,\, \ee^{\sigma q_y l_{\mathcal B}\,
a^{\mathrm{(C)}\dagger}_{\sigma,-\sigma}}\,\, \ee^{-\sigma q_y l_{\mathcal B}\,
a^{\mathrm{(C)}}_{\sigma,-\sigma}}  & \mbox{for} \, \sigma_1= -\sigma_2 \equiv \sigma
\end{array} \right. . \label{eq:spinExp2}
\end{eqnarray}
\endnumparts
Inspection of Eqs.~(\ref{eq:spinExp1})--(\ref{eq:spinExp2}) reveals a very important formal
difference between cases when the interacting particles have equal or opposite spin. For a
pair of particles with the same spin, the interaction only couples to the relative-motion
Landau-level degrees of freedom. In contrast, for particles with opposite spin, the interaction
involves both the relative-motion \emph{and the COM\/} degrees of freedom. This latter
situation is unlike any other encountered previously in the context of fractional QH physics.

With the expressions (\ref{eq:spinExp1})--(\ref{eq:spinExp2}), we are now able to express
the interaction potential for a pair of particles having spin $\sigma_1$ and $\sigma_2$,
respectively, in the basis of COM-angular-momentum and relative-angular-momentum
eigenstates from the lowest Landau level given by
\begin{equation}
|m_\mathrm{C},m_\mathrm{r}\rangle_{\sigma_1\sigma_2} =
\frac{\left[ b^{\mathrm{(C)}\dagger}_{\sigma_1
\sigma_2}\right]^{m_\mathrm{C}}}{\sqrt{m_\mathrm{C}!}} \,
\frac{\left[ b^{\mathrm{(r)}\dagger}_{\sigma_1\sigma_2}
\right]^{m_\mathrm{r}}}{\sqrt{m_\mathrm{r}!}} \, \ket{0}_1\otimes \ket{0}_2 \quad .
\end{equation}
For clarity, the cases where the interacting particles have equal or opposite spin will
be discussed separately.

\subsubsection{Interaction of particles with same spin}

The fact that only the relative-angular-momentum operators enter the expression
(\ref{eq:spinExp1}) for $\sigma_1=\sigma_2\equiv\sigma$ implies that the interaction
matrix is diagonal in COM space. Straightforward calculation yields
\begin{equation}\label{eq:spinlike}
_{\sigma\sigma}\langle m_\mathrm{C}, m_\mathrm{r}|\ee^{i \vek{q}\cdot (\vek{r}_1
- \vek{r}_2)}|m'_\mathrm{C}, m'_\mathrm{r}\rangle_{\sigma\sigma} = \delta_{m_\mathrm{C}
m'_\mathrm{C}}\,\, \ee^{-q^2\,l_{\mathcal B}^2} \,\, {\mathcal M}_{m_\mathrm{r}
m'_\mathrm{r}}(q_\sigma l_{\mathcal B}) \,\, ,
\end{equation}
where
\begin{equation}
{\mathcal M}_{m m'}(\kappa) = \left(\frac{m!}{m'!} \right)^{1/2}(i \kappa)^{m'-m} \,\, 
L^{m'-m}_{m} (|\kappa|^2 )
\end{equation}
in terms of the generalized Laguerre polynomial $L_m^{m'-m}$. Using the result
(\ref{eq:spinlike}) and the relation (\ref{eq:Fourierpot}) for a contact interaction
where $V_{\sigma_1\sigma_2}(\vek{q}) = g_{\sigma_1\sigma_2}$ yields the well-known
expression~\cite{hal90,mac95,coo08}
\begin{eqnarray}\label{eq:usualPP}
_{\sigma\sigma}\langle m_\mathrm{C}, m_\mathrm{r}| V_{\sigma\sigma}(\vek{r}_1 -
\vek{r}_2)  |m'_\mathrm{C}, m'_\mathrm{r} \rangle_{\sigma\sigma} = \frac{g_{\sigma
\sigma}}{4\pi l_{\mathcal B}^2}\,\, \delta_{m_\mathrm{C} m'_\mathrm{C}}\,
\delta_{m_\mathrm{r} m'_\mathrm{r}} \, \delta_{m_\mathrm{r} 0}\quad
\end{eqnarray} 
for the interaction matrix elements. The remarkable result (\ref{eq:usualPP}) underpins
the basic description of fractional-QH physics~\cite{mac95,coo08}. It implies that the
two-particle eigenstates are also eigenstates of COM and relative angular momentum.
Furthermore, the energy spectrum of two particles with spin $\sigma$ from the lowest
Landau level is two-valued: states with $m_\mathrm{r}=0$ and arbitrary $m_\mathrm{C}$
have energy $\hbar\omega_{\mathrm{c}}+g_{\sigma\sigma}/(4\pi l_{\mathcal B}^2)$, and
all other states have energy $\hbar\omega_{\mathrm{c}}$.

\subsubsection{Interaction of particles with opposite spin}

Using (\ref{eq:spinExp1}) for the case $\sigma_1=-\sigma_2\equiv\sigma$, we find 
\begin{eqnarray}
&& _{\sigma,-\sigma}\langle m_\mathrm{C}, m_\mathrm{r}|\ee^{i \vek{q}\cdot
(\vek{r}_1 - \vek{r}_2)}|m'_\mathrm{C}, m'_\mathrm{r}\rangle_{\sigma,-\sigma} = \ee^{-q^2\,
l_{\mathcal B}^2} \nonumber \\ && \hspace{4cm} \times \,\, (i\sigma)^{m'_\mathrm{C} -
m_\mathrm{C}}\,\, {\mathcal M}_{m_\mathrm{r} m'_\mathrm{r}}(q_x l_{\mathcal B}) \,\,
{\mathcal M}_{m_\mathrm{C} m'_\mathrm{C}}(q_y l_{\mathcal B}) \,\, .
\end{eqnarray}
The contact-interaction matrix element for opposite-spin particles is then calculated as
\begin{eqnarray}\label{eq:oppSpin}
 && _{\sigma,-\sigma}\langle m_\mathrm{C}, m_\mathrm{r}| V_{\sigma,-\sigma}(\vek{r}_1
- \vek{r}_2)  |m'_\mathrm{C}, m'_\mathrm{r} \rangle_{\sigma,-\sigma} \nonumber \\ &&
\hspace{6cm} = \frac{g_{+-}}{\left(2\pi l_{\mathcal B}\right)^2} \,\, i^{m'_\mathrm{r} -
m_\mathrm{r}}\,\, {\mathcal R}_{m_\mathrm{r} m'_\mathrm{r}}\,\, {\mathcal R}_{m_\mathrm{C}
m'_\mathrm{C}} \,\, ,
\end{eqnarray}
with
\begin{equation}
{\mathcal R}_{m m'} = \frac{[1+(-1)^{m'-m}]}{2\sqrt{m!\, m'!}} \,\, \Gamma\left(
\frac{m+m'+1}{2}\right)
\end{equation}
in terms of the Euler Gamma function $\Gamma(x)$. Thus we find that the interaction
matrix for two particles from the lowest Landau level with opposite spin is nondiagonal in
the COM-angular-momentum and relative-angular-momentum spaces. This is markedly
different from the case of same-spin particles.

\begin{figure}
\begin{center}
\includegraphics[width=0.48\columnwidth]{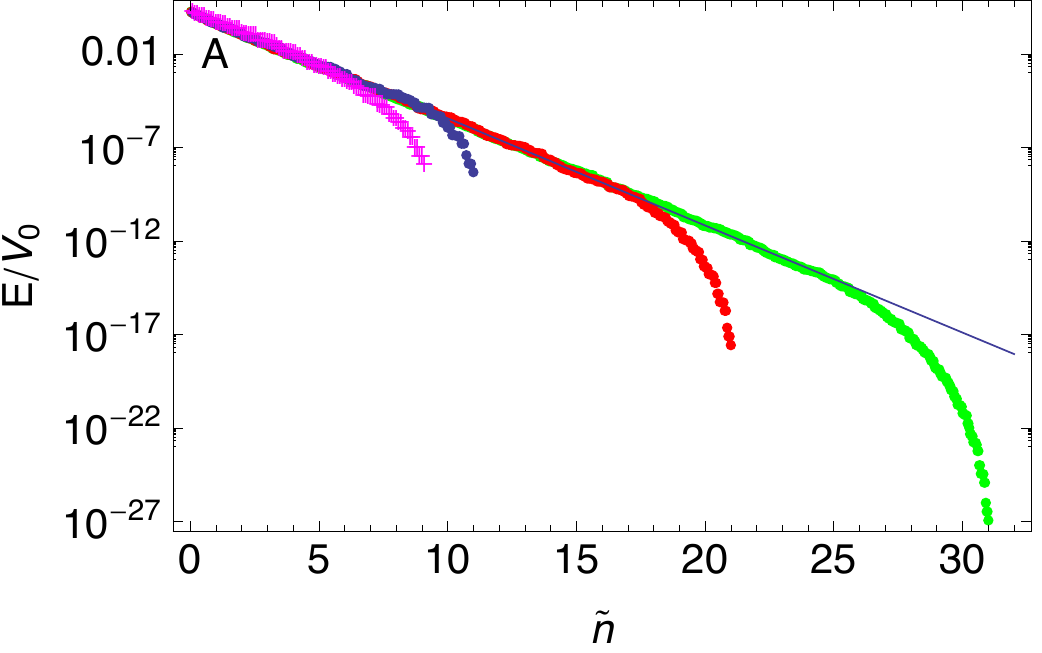}
\includegraphics[width=0.5\columnwidth]{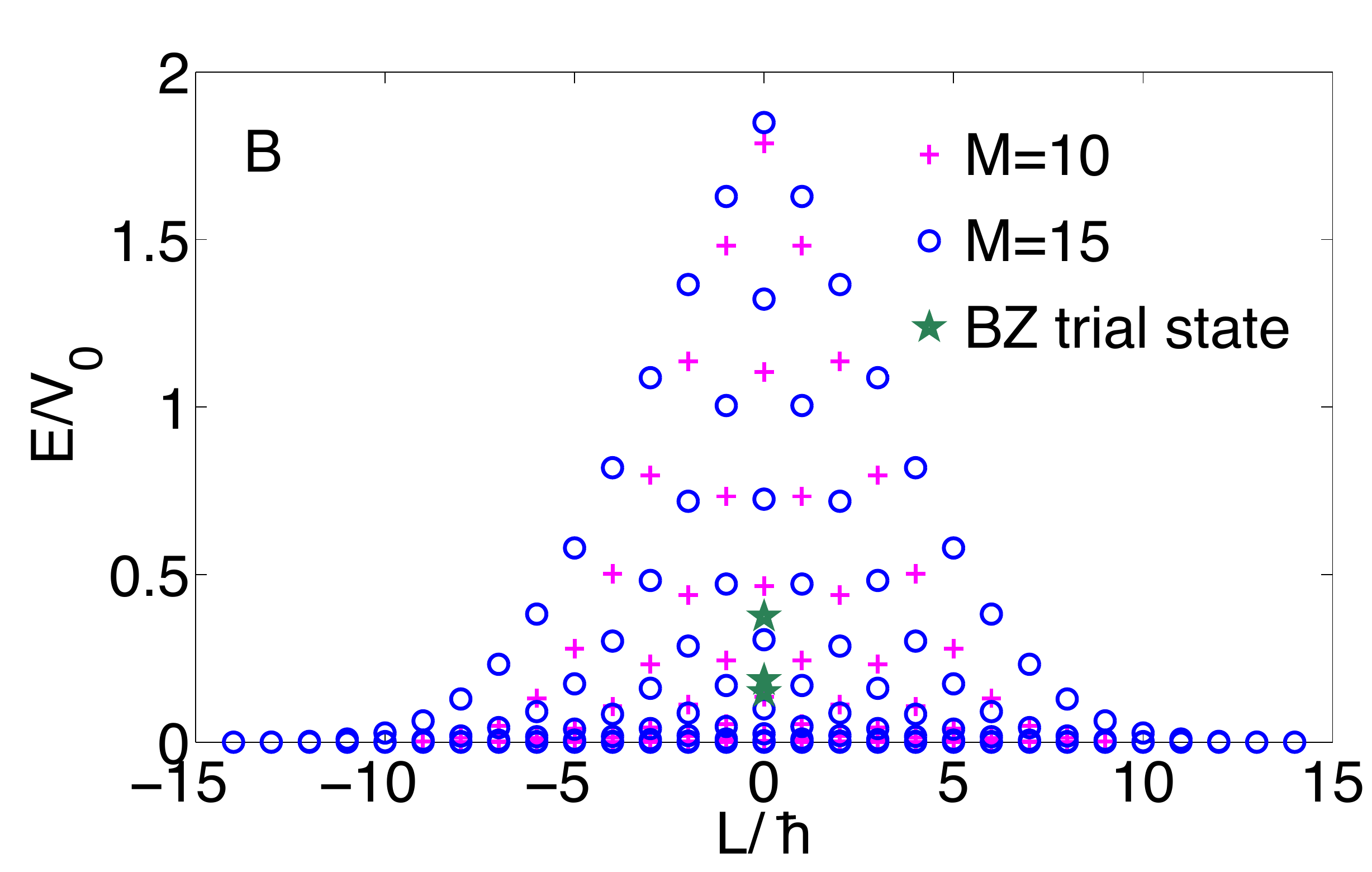}\\ [0.2cm]
\includegraphics[width=0.48\columnwidth]{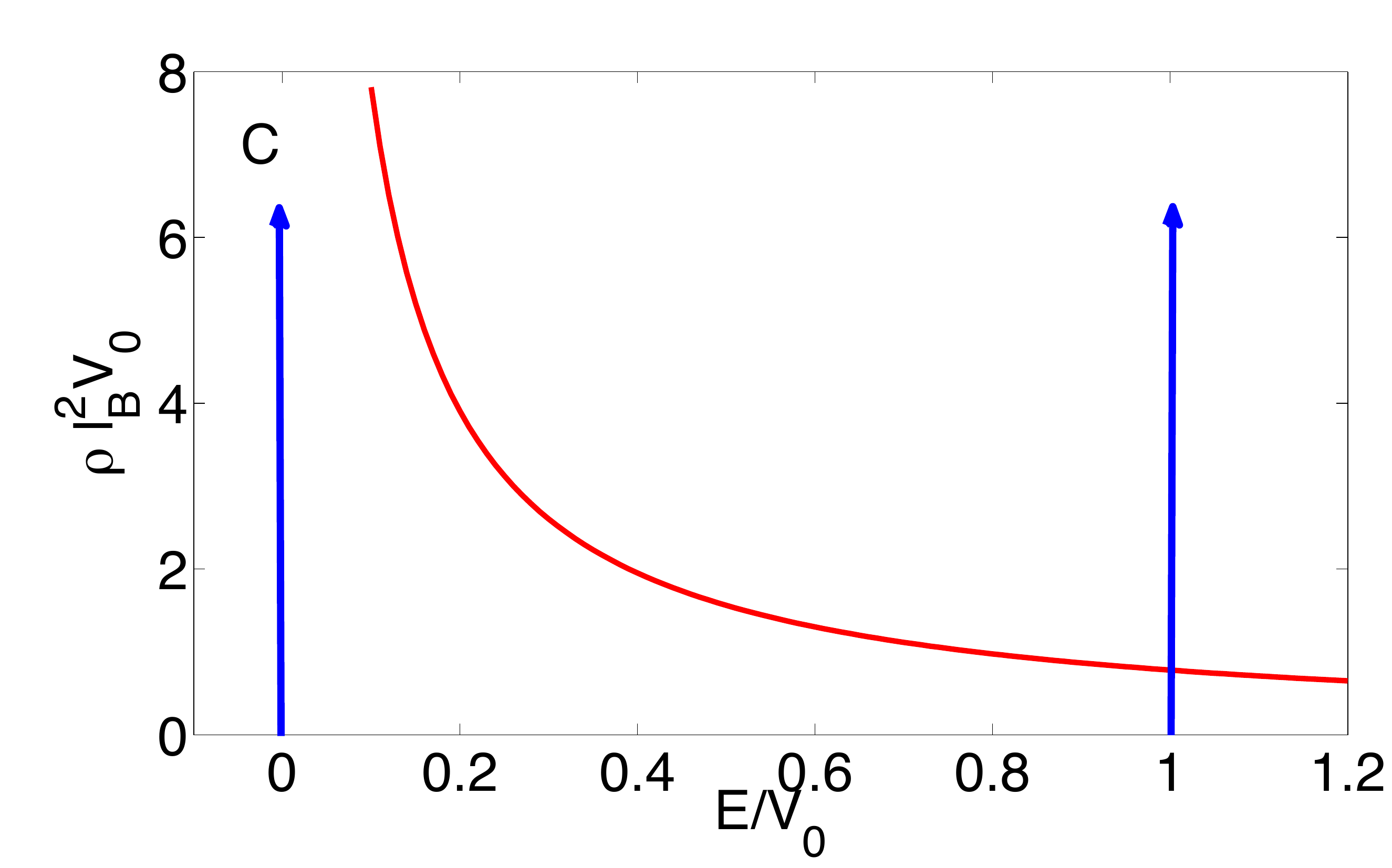}
\end{center}
\caption{\label{2ParticlesTheo}
Panel A: Eigenvalues $E$ of the opposite-spin two-particle interaction matrix [cf.\
Eq.~(\ref{eq:oppSpin})] in units of  $V_0\equiv g_{+-}/(4\pi l^2_{\mathcal B})$, 
sorted by magnitude. Data are shown for various values of the angular-momentum
cutoff $m_\mathrm{max}=10$ (blue), 20 (red), 30 (green), and $\tilde{n} = n/
(m_\mathrm{max}+1)$. The straight line is a plot of $E= 0.3\,V_0\exp(-\alpha \tilde{n})$
with $\alpha= 1.28$.  Panel B: Energy spectrum obtained for a system of two particles
with opposite spin by exact diagonalization. A finite system size is imposed by
limiting the number of modes available in angular-momentum space for each particle
to $\mathcal{M}$. Note the $\mathcal{M}$-dependence of the obtained values. The
data for $\mathcal{M}=10$ are also shown as the magenta data points in panel A and
exhibit excellent agreement with the power-law-type distribution predicted from the
solution in COM and relative angular-momentum space. Green stars show the energy
calculated for two-particle versions of trial states~\cite{ber06a} $\psi_{+-}(\vek{r}_1,
\vek{r}_2) \propto (z_1+z_2^\ast)^{m_\mathrm{C}}(z_1-z_2^\ast)^{m_\mathrm{r}}$
with $m_\mathrm{C}=0$ and $m_\mathrm{r} = 2$, $9$, $14$. Higher $m_\mathrm{r}$
is observed to correlate with lower energy, but there are many states even lower in
energy than the trial state with largest $m_\mathrm{r}$ that is compatible with the finite
systems size. Panel C: Comparison of two-particle densities of states for same-spin
case (blue arrows indicating delta functions) and for opposite-spin case (red curve).}
\end{figure}

Straightforward diagonalization of the matrix (\ref{eq:oppSpin}) yields the two-particle
eigenenergies $E_n$ when both particles have opposite spin. Figure~\ref{2ParticlesTheo}A
shows a logarithmic plot of the $E_n$, ordered by decreasing magnitude, for different
values $m_{\mathrm{max}}$ of the cut-off value for COM and relative angular momentum.
We observe an exponential dependence of the sorted eigenvalues as a function of the scaled
index $\tilde{n}=n/(m_\mathrm{max}+1)$, which translates into a power-law density of
states
\begin{equation} \label{eq:dos}
\rho(E) = \left| \frac{\Delta n}{\Delta A \Delta E}\right| \approx \frac{1}{\alpha
l_{\mathcal B}^{2} E} \quad .
\end{equation}
Here $\Delta A = (m_\mathrm{max}+1)  l_{\mathcal B}^{2}$ is the area corresponding to
the cut-off in COM and relative angular momentum, and $\alpha \approx 1.28$ has been
determined numerically. The numerical data deviate from Eq.~(\ref{eq:dos}) close to the
maximum energy $g_{+-}/({2\pi l_{\mathcal B}^{2}})$, where the density of states reaches
zero, and for small energy where it becomes cutoff dependent. None of the individual
eigenvalues is strictly independent of the cutoff, which indicates that there are no
compact eigenstates. Figure~\ref{2ParticlesTheo}C illustrates the different density-of-states
behavior for interacting two-particle systems for the two cases of particles having the
same and opposite spin, respectively.

It is also useful to look at the distribution of eigenvalues over total angular momentum.
We do this with a different numerical scheme using exact diagonalization of the two-particle
Hilbert space on a disk, as it preserves the $z$ component of angular momentum as a
good quantum number. (Details are given in the following section.) The spectrum for
$N_+ = N_- =1$ is shown in Fig.~\ref{2ParticlesTheo}B. Note the dependence of the
eigenvalues on the systems size (ie, the cut-off in angular momentum of available
Landau-level states). Any systematic difference between the results given in
Figs.~\ref{2ParticlesTheo}A and \ref{2ParticlesTheo}B is probably at least in part due
to the fact that the representation using the COM and relative angular-momentum basis
assumes an infinite number of single-particle angular-momentum modes to be available
to the particles. Nevertheless, when the energy eigenvalues obtained for the finite system
are plotted alongside the results for the analytical model (see magenta data points in
Fig.~\ref{2ParticlesTheo}A), both are seen to exhibit the same exponential behavior. For
comparision, the energies calculated for proposed trial states~\cite{ber06a} are also
shown in Fig.~\ref{2ParticlesTheo}B as green stars.

\section{Few-particle systems studied by numerically exact diagonalization
\label{sec:numFew}}

While the interacting two-particle problem has lent itself to analytical study, the
behavior of systems with three or more interacting particles either requires
approximate, e.g., variational, approaches or must be done numerically. As the
complications encountered already for the case of two interacting particles with
opposite spin stymie progress for the variational option, we follow the numerical
route here.

\subsection{Formalism and results for few-particle eigen-energy spectra}

We start by representing the Schr\"odinger field operator for a particle at position
$\rr$ with spin $\sigma$ projected onto the lowest spin-related Landau level,
\begin{equation}
\hat{\Psi}_\sigma^{(\mathrm{LLL})}(\rr) = \sum_{m\ge 0} \phi^{(\sigma)}_{0, m}(\rr)
\,\, \hat{c}_{\sigma m} \quad ,
\end{equation}
where $\hat{c}^{\dagger}_{\sigma m}$ creates a particle in component $\sigma$
with angular momentum $\sigma m$ in the state $ \phi^{(\sigma)}_{0, m}(\rr)\equiv
\bra{\rr}\left(b^\dagger_\sigma\right)^m /\sqrt{m!} \ket{0}$.
Substituting this into Eq.~(\ref{eq:model_int}), we get
\begin{equation}
{\mathcal H}_{\mathrm{int}}^{(\mathrm{LLL})} = \sum_{\sigma \sigma'}\sum_{\{m\}}
\Lambda_{\{m\}}^{(\sigma, \sigma')}\hat{c}^\dagger_{\sigma m_1}
\hat{c}^\dagger_{\sigma' m_2} \hat{c}_{\sigma' m_3} \hat{c}_{\sigma m_4},
\end{equation}
where
\begin{equation}
\Lambda_{\{m\}}^{(\sigma,\sigma')}=\frac{g_{\sigma\sigma'}}{2}\int d^2 r \,\,  \left[
\phi^{(\sigma)}_{0, m_1}(\vek{r})\right]^\ast \left[\phi^{(\sigma')}_{0, m_2}(\vek{r}) \right]^\ast
\phi^{(\sigma')}_{0,m_3}(\vek{r})\, \phi^{(\sigma)}_{0, m_4}(\vek{r}) \quad .
\end{equation}
For same-spin particles, i.e., $\sigma'=\sigma$, we obtain
\begin{equation}\label{eq:sameInt}
\Lambda_{\{m\}}^{(\sigma, \sigma)} = \frac{g_{\sigma\sigma}}{8\pi l^2_{\mathcal B}}
\frac{(m_1+m_2)!\delta_{m_1+m_2,m_3+m_4}}{2^{m_1+m_2}\sqrt{m_1!m_2!m_3!m_4!}}
\quad .
\end{equation}
In contrast, for the matrix element involving opposite-spin particles ($\sigma =
-\sigma^\prime$), we find
\begin{equation}\label{eq:oppositeInt}
\Lambda_{\{m\}}^{(\sigma,-\sigma)} = \frac{g_{\sigma,-\sigma}}{8\pi l^2_{\mathcal B}}
\frac{(m_1+m_3)!\delta_{m_1+m_3,m_2+m_4}}{2^{m_1+m_3}\sqrt{m_1!m_2!m_3!m_4!}}
\quad .
\end{equation}
The way indices are distributed in the arguments of the $\delta$-functions in
Eqs.~(\ref{eq:sameInt}) and (\ref{eq:oppositeInt}) implies that the system's total angular
momentum $L \equiv \sum_j L_{z j}$ [cf.\ Eq.~(\ref{eq:angMom}) for the
definition of $L_z$] is a conserved quantity in the presence of interactions.

Cold-atom systems are usually studied while trapped by an external potential of tunable
strength. To model this situation, we introduce the second-quantized form of a
parabolic potential in the representation of lowest-Landau-level states,
\begin{equation}
{\mathcal H}_0^{(\mathrm{LLL})} = \sum_{\sigma, m} \alpha (m+1) \,
\hat{c}^{\dagger}_{\sigma m} \hat{c}_{\sigma m} \quad ,
\end{equation}
where $\alpha = M \Omega^2 l_{\mathcal B}^2$ in terms of the harmonic-trap
frequency $\Omega$. Switching on the trap will lift degeneracies of few-particle states
and serve to identify the most compact ground states of our systems of interest.

\begin{figure}[t]
\begin{center}
\includegraphics[width=12cm,height=9cm]{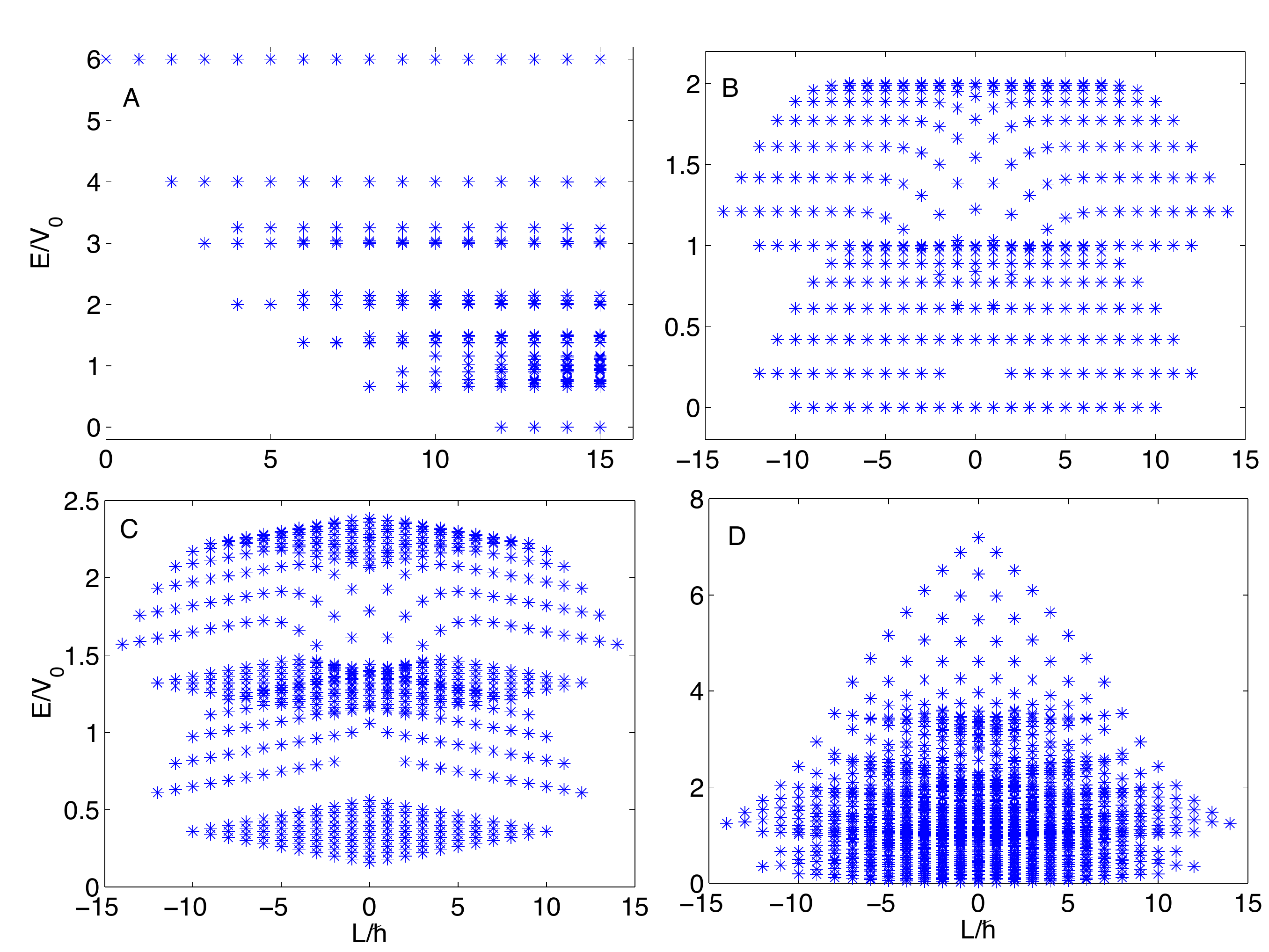}
\end{center}
\caption{\label{fig:n2n2spec}
Spectrum for various four-particle systems (i.e., $N_+ + N_- = 4$). Energies
are given in units of the intra-species Haldane-pseudopotential energy scale
$V_0 = g_{++}/(4\pi l_{\mathcal B}^2)$. A: Single-component system with
$N_+=4$, $N_-=0$. The four-particle Laughlin state is the zero-energy state
with the smallest total angular momentum $L=12$. B: System with $N_+ =
N_- = 2$ and $g_{++} = g_{--} \ne 0$, $g_{+-} = 0$ (no interspecies interaction).
C: Same situation as for B but with a finite trapping potential ($\alpha=0.02$)
switched on in addition, revealing the energy degeneracies in B. The
lowest-energy state is a superposition of two-particle Laughlin states in each
component. D: Same situation as for B but with finite interspecies interaction
$g_{+-}=g_{++}$ in addition.}
\end{figure}

We calculate the few-particle energy spectra and associated eigenstates for
${\mathcal H}_0^{(\mathrm{LLL})} + {\mathcal H}_{\mathrm{int}}^{(\mathrm{LLL})}$ in
the Fock basis of lowest-Landau-level states for the two spin components. We focus
here on the case of bosonic particles to be directly applicable to currently studied
ultra-cold atom systems, but our general conclusions apply to systems of fermionic
particles as well. Results obtained for systems with $N_+ + N_- = 4$ are shown in
Fig.~\ref{fig:n2n2spec}. Panel A shows the situation where only particles from a single
component are present, which is analogous to the previously
considered  case of spinless bosons \cite{coo99,wil00,vie00,vie08}. The zero-energy
state at lowest total angular momentum has $|L|=N(N-1)$ and
corresponds to the filling-factor-$1/2$ Laughlin state~\cite{coo08,vie08}. Zero-energy
eigenstates at higher magnitudes of total angular momentum correspond to edge
excitations of the Laughlin state~\cite{mac95}. The $L=0$ state has an
energy of $V_0 N(N-1)/2$, where $V_0\equiv g_{++}/(4\pi l^2_{\mathcal B})$.

When particles occupy states in both components, the situation becomes complex.
Without interaction between the different spin species, states of each component will
be the ones that are obtained by diagonalising the interacting Hamiltonian within that
component. The entire system is then essentially an independent superposition of
eigenstates for the individual spin species. However, in contrast to ordinary multi-component
QH states discussed, e.g., in Refs.~\cite{hal83a,rei02,rei04}, the total angular momenta
for states from different components have opposite sign. Therefore, e.g., the combination
of Laughlin states in each component with the same number of particles has zero total
angular momentum. This case is illustrated in Fig.~\ref{fig:n2n2spec}B. However, the
superpositions of edge excitations with same magnitude of excess angular momentum
for the opposite-spin Laughlin states will also be zero-energy, zero-angular-momentum
eigenstates. To reveal the associated degeneracies of the spectrum shown in
Fig.~\ref{fig:n2n2spec}B, we obtained the energy eigenvalues in the presence of a
parabolic confinement. See Fig.~\ref{fig:n2n2spec}C. Notice the band of low-lying energy
levels separated by a gap from higher-energy states. The lowest-energy
$L=0$ state is the superposition of the two-particle Laughlin states for
the two spin species. The other states in the low-energy band correspond to edge
excitations of this configuration.

Figure~\ref{fig:n2n2spec}D illustrates the dramatic effect of interactions between
opposite-spin particles. The spectrum seen there has to be compared with that
given in panel~B where only particles with the same spin interact. Note the disappearance
of energy gaps and accumulation of states at low energy, reflecting the characteristic
features of the opposite-spin two-particle interaction spectrum shown in
Fig.~\ref{2ParticlesTheo}B. Clearly, the system is not incompressible anymore, and
no QH-related physics can be expected to occur.

\begin{figure}[t]
\begin{center}
\includegraphics[width=12cm,height=9cm]{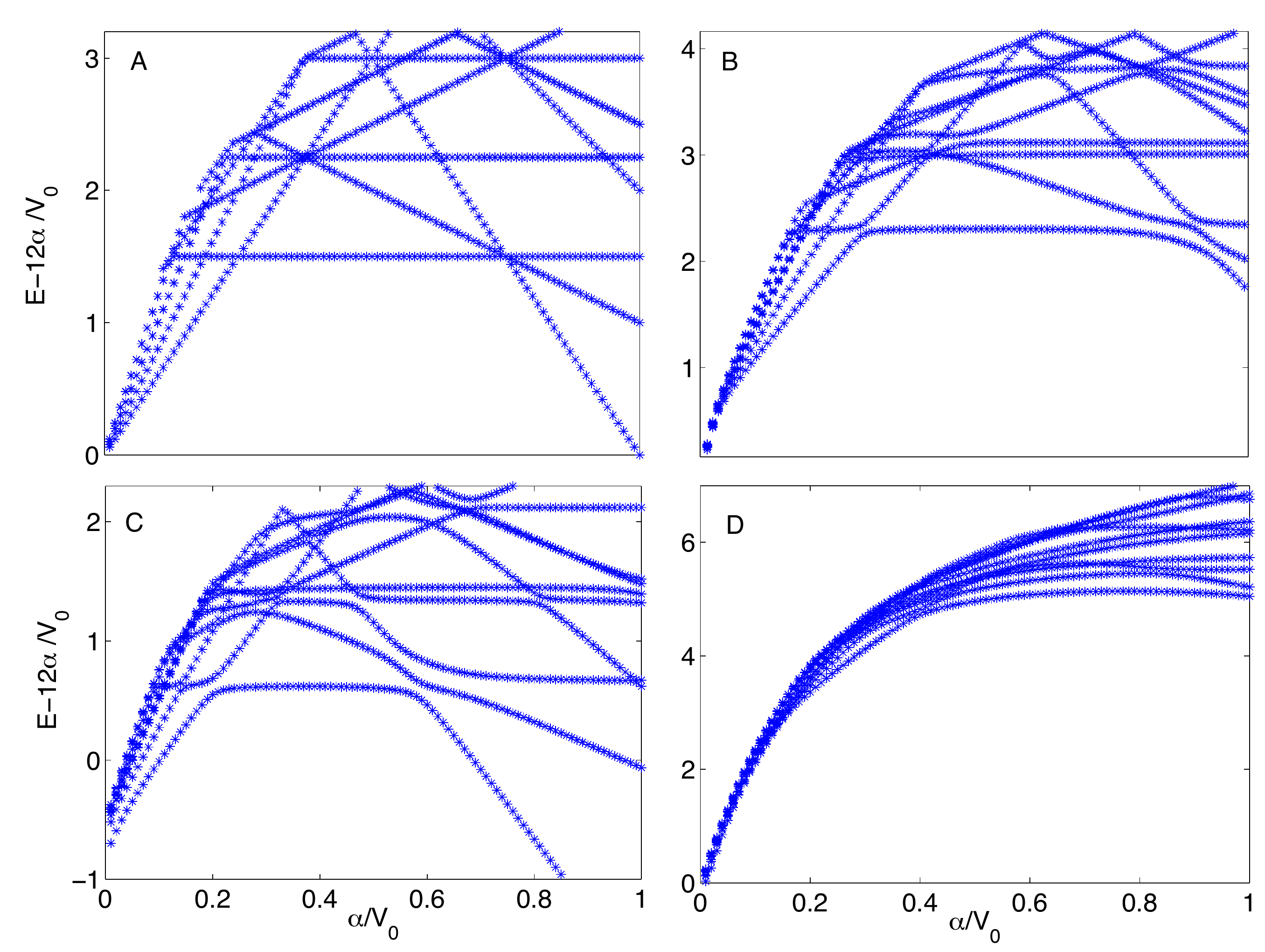}
\end{center}
\caption{\label{fig:energy3}
Low-lying energy levels for a system with $N_+ = N_- = 3$ in the sector of total
angular momentum $L=0$. In the calculation, lowest-Landau-level states with $m\le 18$
have been included. Different panels correspond to different interspecies-interaction
strengths. A: No inter-species interactions ($g_{+-} = 0$). A finite trapping potential
lifts the energy degeneracies seen at $\alpha=0$ and singles out a unique lowest-energy
state. For small $\alpha$, the latter turns out to be the superposition of Laughlin states
for each individual component. At $\alpha=0.2$ it becomes an incompressible state with a
single Laughlin quasi-particle in each component. Finally, at $\alpha=0.8$ both components
are Bose-condensed in the lowest Landau level. Modest interspecies-interaction strengths
($g_{\sigma\bar{\sigma}}=0.2\, V_0$ in panel B and $g_{\sigma\bar{\sigma}}=-0.2\, V_0$
in panel C) cause avoided crossings but preserve the incompressible nature of the states
seen in panel A. A stronger interspecies interaction ($g_{+-} = V_0$ in panel D) washes out
that picture completely.}
\end{figure}

In Figure~\ref{fig:energy3}, the interplay between interactions and confinement is
elucidated. Panels~A--D show the evolution of low-lying few-particle eigenstates as
the confinement strength is varied for situations with different magnitude of interaction
strength between opposite-spin particles. Panel~A corresponds to the case with
$g_{+-} = 0$. Due to the occurrence of level crossings, the character of the lowest-energy
(ground) state is found to be different for regimes associated with weak, intermediate,
and strong confinement. Analogous behavior has been discussed previously for
ordinary (spinless) few-boson fractional QH systems~\cite{pop04}. In our case
depicted in Fig.~\ref{fig:energy3}A, the ground state in the weak-confinement regime
corresponds to a superposition of three-particle Laughlin states for filling factor $1/2$
in the individual pseudospin components. After the first level crossing, each component
turns out to be in the Laughlin-quasiparticle state~\cite{pop04} and, after another
level crossing, each spin component has its three particles occupying the lowest
state defined by the parabolic confinement potential. (Our description of the ground
states found in the three different regimes is supported by the analysis of real-space
density and angular-momentum distribution functions. See the following subsection
for details.)

Switching on interactions between opposite-spin particles turns crossings into
anti-crossings. Figures.~\ref{fig:energy3}B and \ref{fig:energy3}C depict situations
where interactions between same-spin particles are still dominant. The existence
of anticrossings enables smooth transitions between the different ground states
that would not be possible in the case of simple crossings as seen, e.g., in
panel~A. Independent tuning of interactions between opposite-spin particles can
therefore be used to enable engineering of quantum many-particle states in ways
not anticipated in previous work~\cite{pop04}.

Strong interactions between opposite-spin particles are again seen to fundamentally
alter the character of the system's ground and excited states. In Fig.~\ref{fig:energy3}D,
the strengths of interactions between same-spin and opposite-spin particles are equal.
The variation of few-particle states as a function of confinement strength is seen to
be almost uniform, again pointing to the loss of distinctiveness for few-particle states
in the presence of inter-species interactions. Furthermore, energy differences between
low-lying states are much reduced as compared to the situation depicted in panel A
of the same figure, which is a reflection of the unusual distribution of energy eigenvalues
found for the interacting opposite-spin two-particle system.

\subsection{Results for physical properties of the few-particle ground state}

The one-particle density profiles in coordinate space and in angular-momentum
space are useful quantities to enable greater understanding of the properties of
specific many-body quantum states~\cite{coo05,dou11}. In the basis of
lowest-Landau-level states from the two spin components, the single-particle
density matrix of a many-particle state $\ket{\Phi}$ has matrix elements
\begin{equation}
\rho_{\sigma m, \sigma' m'}= \bra{\Phi} \hat c^\dagger_{\sigma m} \hat
c_{\sigma' m'}\, \ket{\Phi} \quad .
\end{equation}
In terms of this quantity, we can define the angular-momentum distribution for
each spin component,
\begin{equation}
\rho^{(\sigma)}_{m} = \rho_{\sigma m, \sigma m} \quad ,
\end{equation}
and also the spin-resolved single-particle density profile in real space,
\begin{equation}
n^{(\sigma)}(\rr) = \sum_{m, m'} \rho_{\sigma m, \sigma m'} \left[
\phi^{(\sigma)}_{0, m}(\rr) \right]^\ast \phi^{(\sigma)}_{0, m'}(\rr) \quad .
\end{equation}
In the following, we focus on the properties of the lowest-energy (ground) state
in the different regimes associated with small, intermediate, and strong
confinement strength for the systems whose energy spectra are shown in
Fig.~\ref{fig:energy3}.

\begin{figure}[t]
\begin{center}
\includegraphics[width=.4\columnwidth]{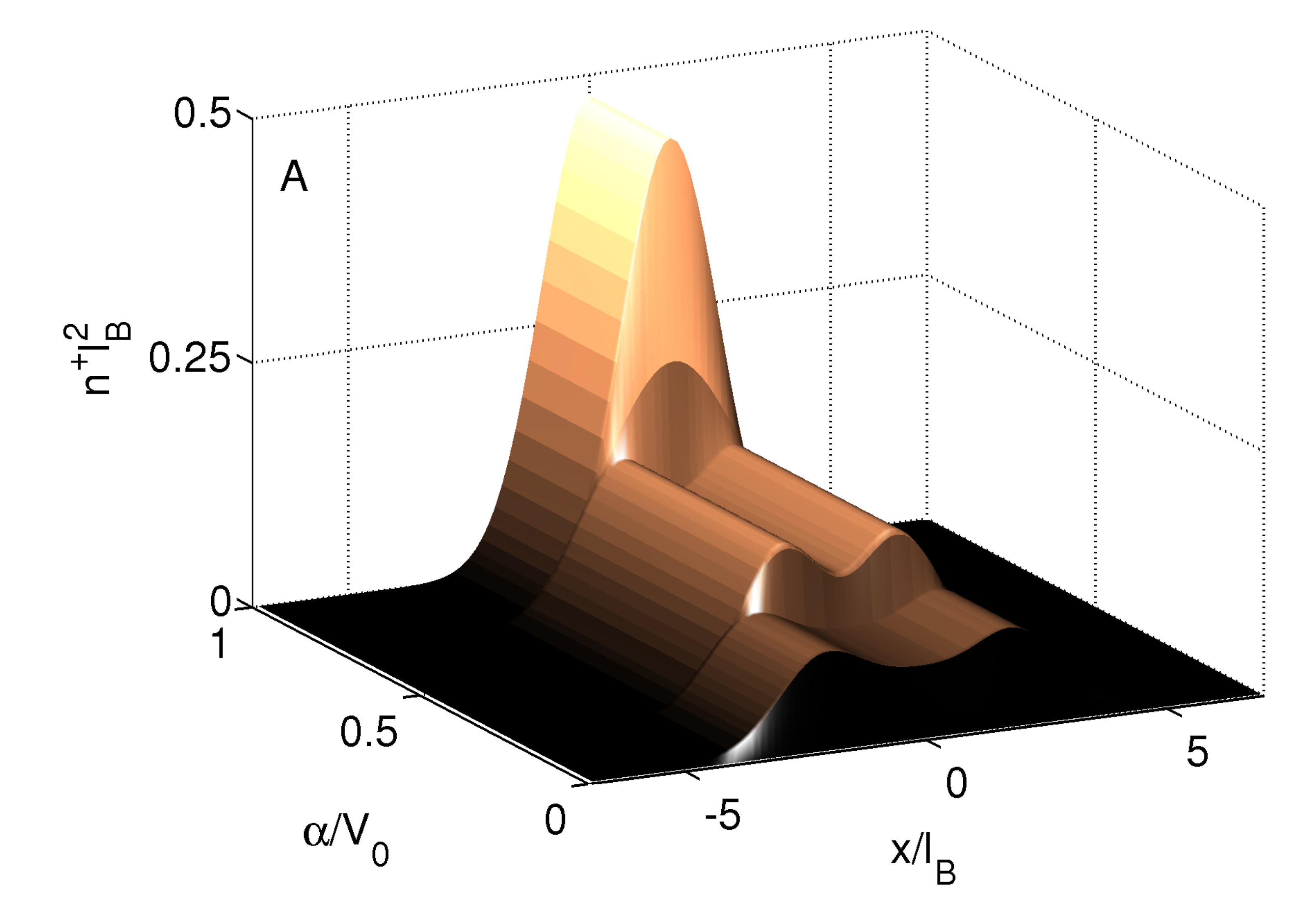}\hspace{0.5cm}
\includegraphics[width=.4\columnwidth]{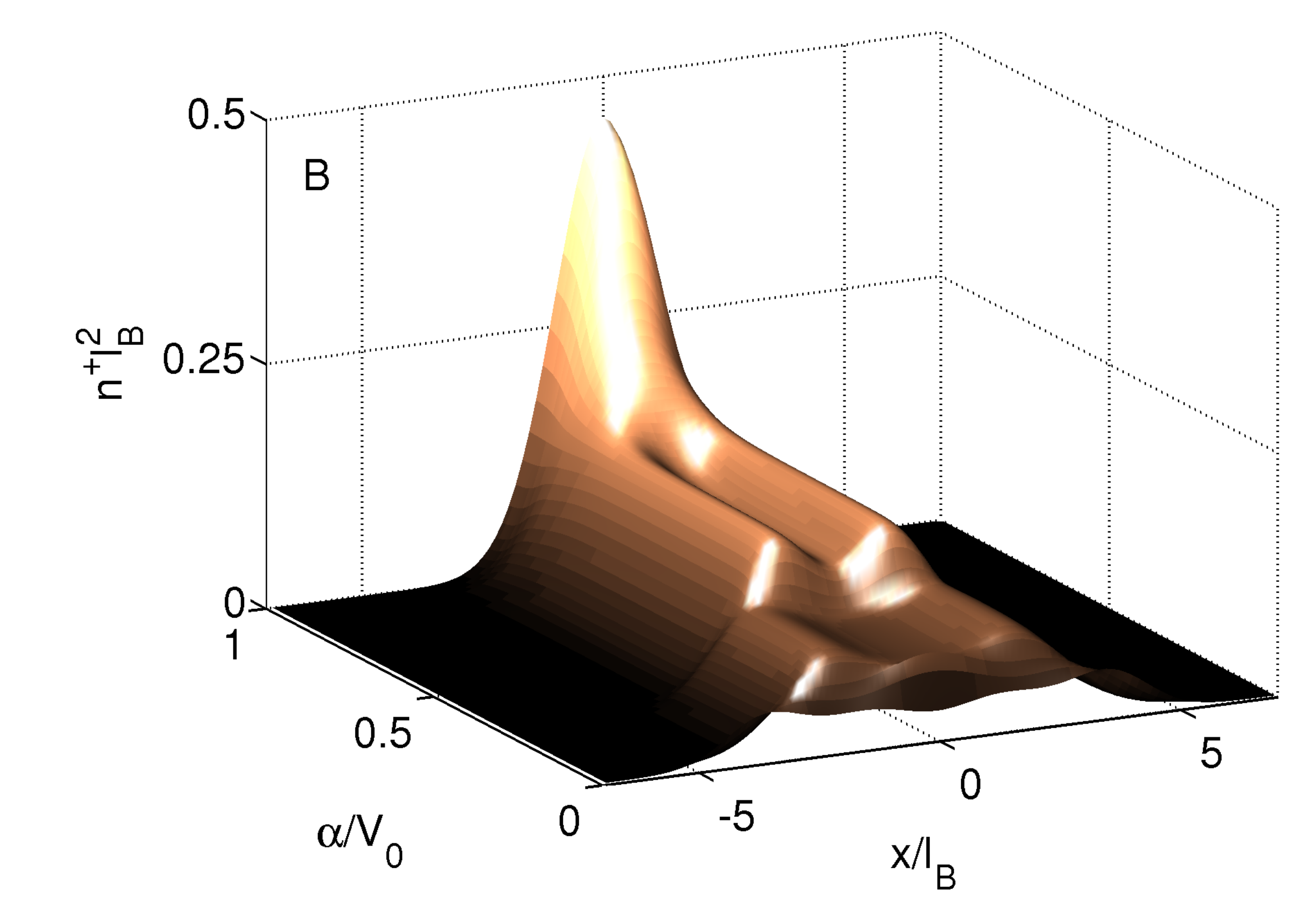}\\
\includegraphics[width=.4\columnwidth]{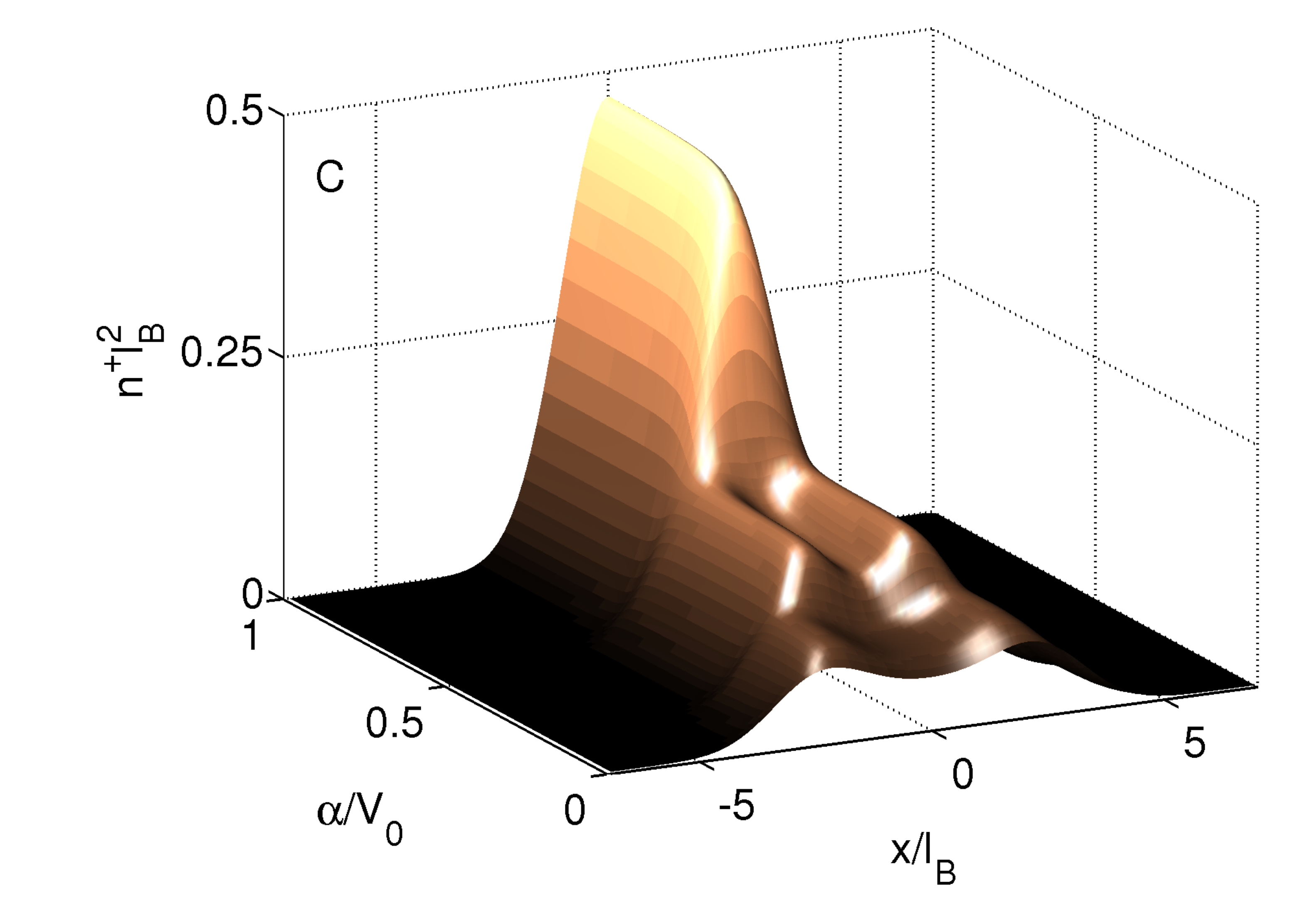}\hspace{0.5cm}
\includegraphics[width=.4\columnwidth]{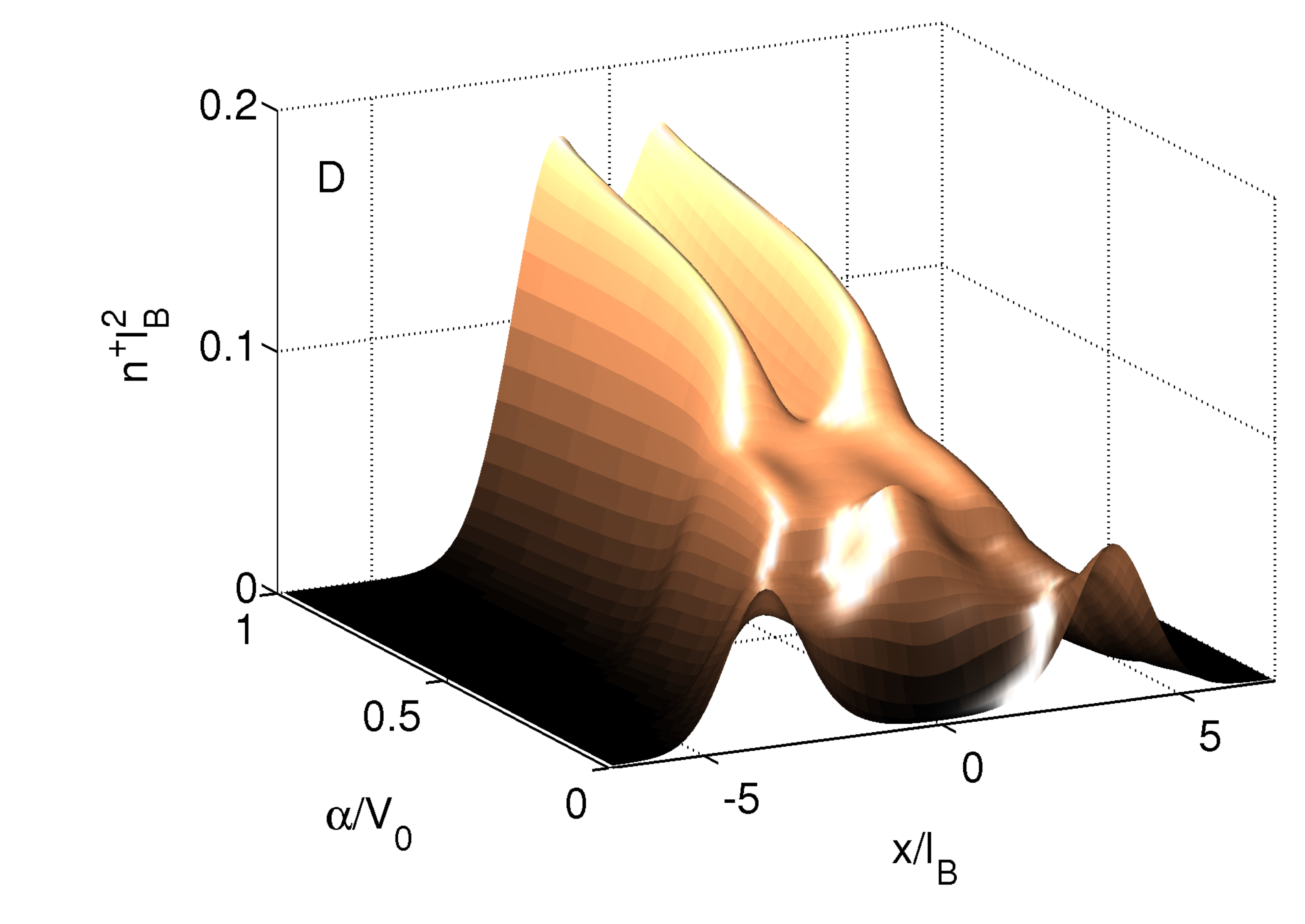}
\end{center}
\caption{\label{fig:densities3}
Cross-sectional pseudo-spin $+$ density profiles of the few-particle ground state
associated with the lowest-lying energy level shown in the corresponding panels
A--D of Fig.~\ref{fig:energy3}, aggregated as a function of the confinement-potential
strength $\alpha$. In panel A (only particles with same spin interact), sharp
transitions occur between the FQH (Laughlin) state in the regime of small $\alpha$,
a Laughlin-quasiparticle-type state for intermediate $\alpha$, and the Gaussian
Bose-Einstein-condensed state at high $\alpha$. For moderate interaction strength
between opposite-spin components (repulsive in panel B, attractive in panel C),
transitions become smooth crossovers associated with anticrossings in
Fig.~\ref{fig:energy3}. Stronger interactions strengths between the spin components
significantly change the character of the few-particle state at small $\alpha$ (panel D).}
\end{figure}

Figure~\ref{fig:densities3} shows the real-space profile of $n^{(+)}(\rr)$ across a
diameter of the disk-shaped three-particle systems associated with the ground-state
levels shown in Fig.~\ref{fig:energy3}. In the absence of interactions between
opposite-spin particles, each component realizes correlated few-particle states
of the type that have been found in previous work~\cite{pop04}. The few-particle
filling-factor-1/2 FQH state is the ground state for a weak confinement potential.
Increasing the trapping-potential strength favors more compact correlated states,
hence, at a critical value of $\alpha$, a transition occurs to a three-particle version
of the Laughlin-quasiparticle excited state. At even higher $\alpha$, the system
transitions to the Gaussian Bose-Einstein-condensate state. The sharpness of
the transitions reflects the existence of level crossings in Fig.~\ref{fig:energy3}A.
Practically, simple variation of $\alpha$ would not lead to any such transitions
because there is no mechanism for the system to switch between different
many-particle states. To make adiabatic passage between different many-particle
states possible, some symmetry of the system needs to broken, and previous work
has proposed scenarios for achieving this in the single-component case~\cite{pop04}.

\begin{figure}
\begin{center}
\includegraphics[width=.4\columnwidth]{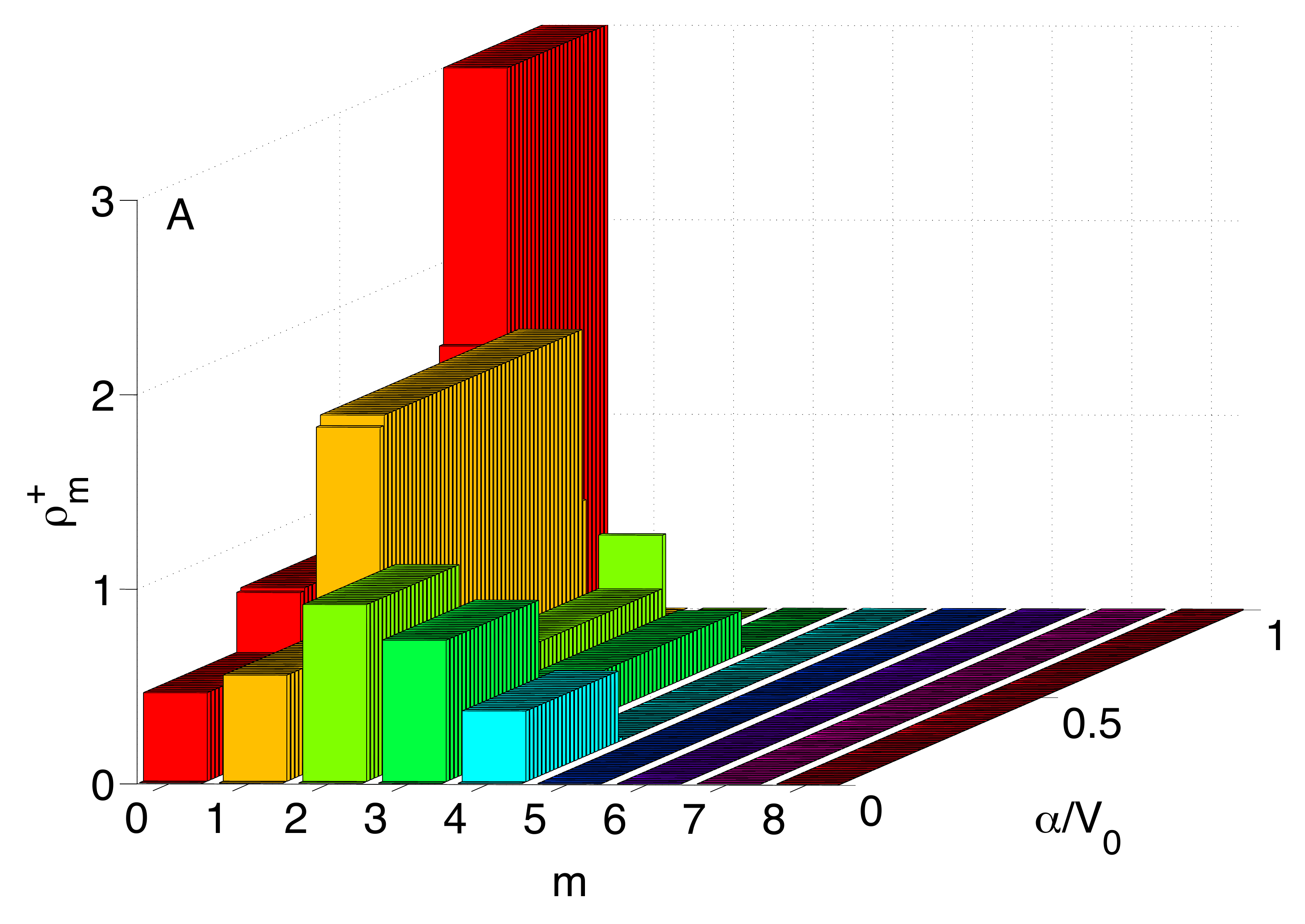}\hspace{0.5cm}
\includegraphics[width=.4\columnwidth]{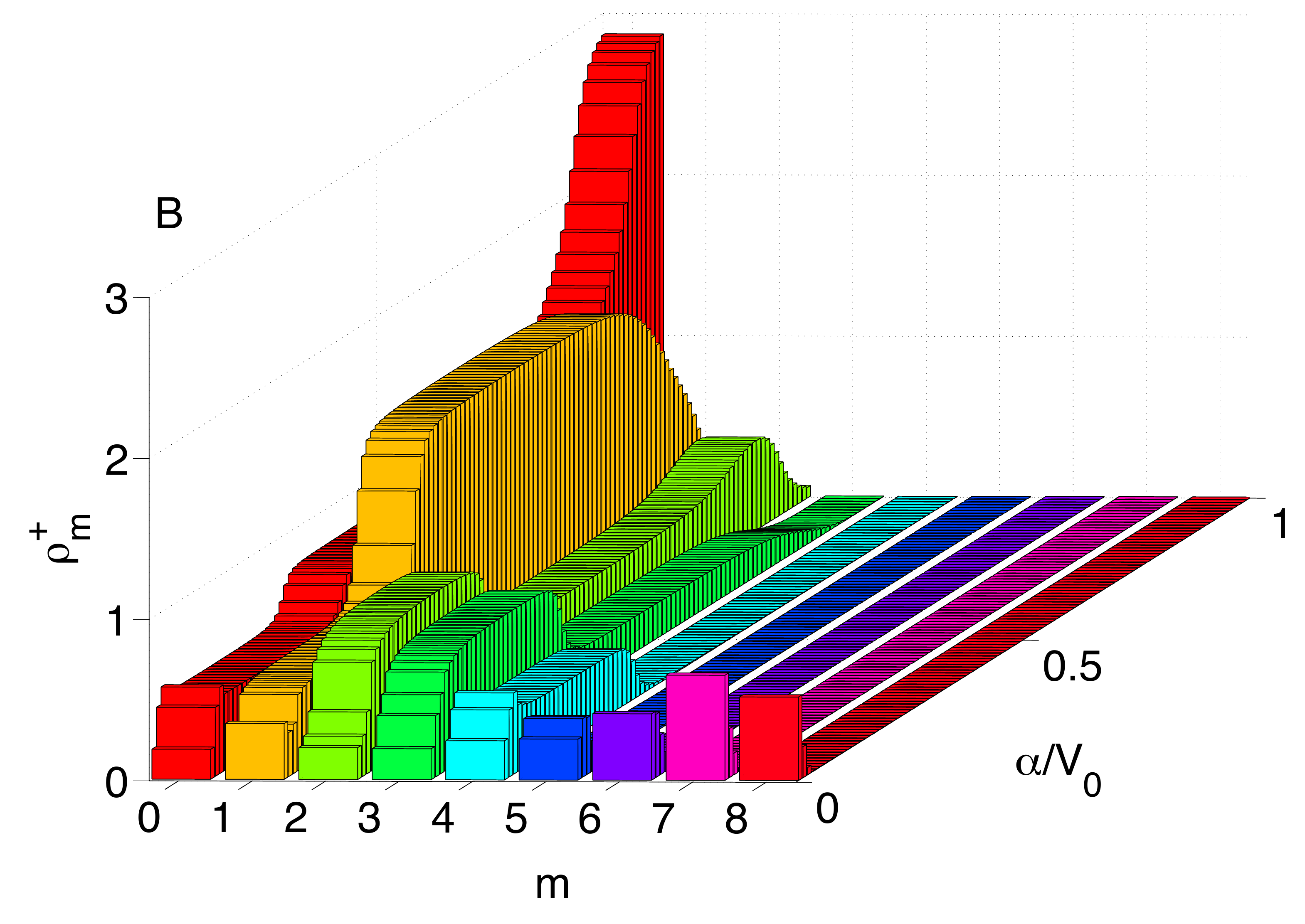}\\
\includegraphics[width=.4\columnwidth]{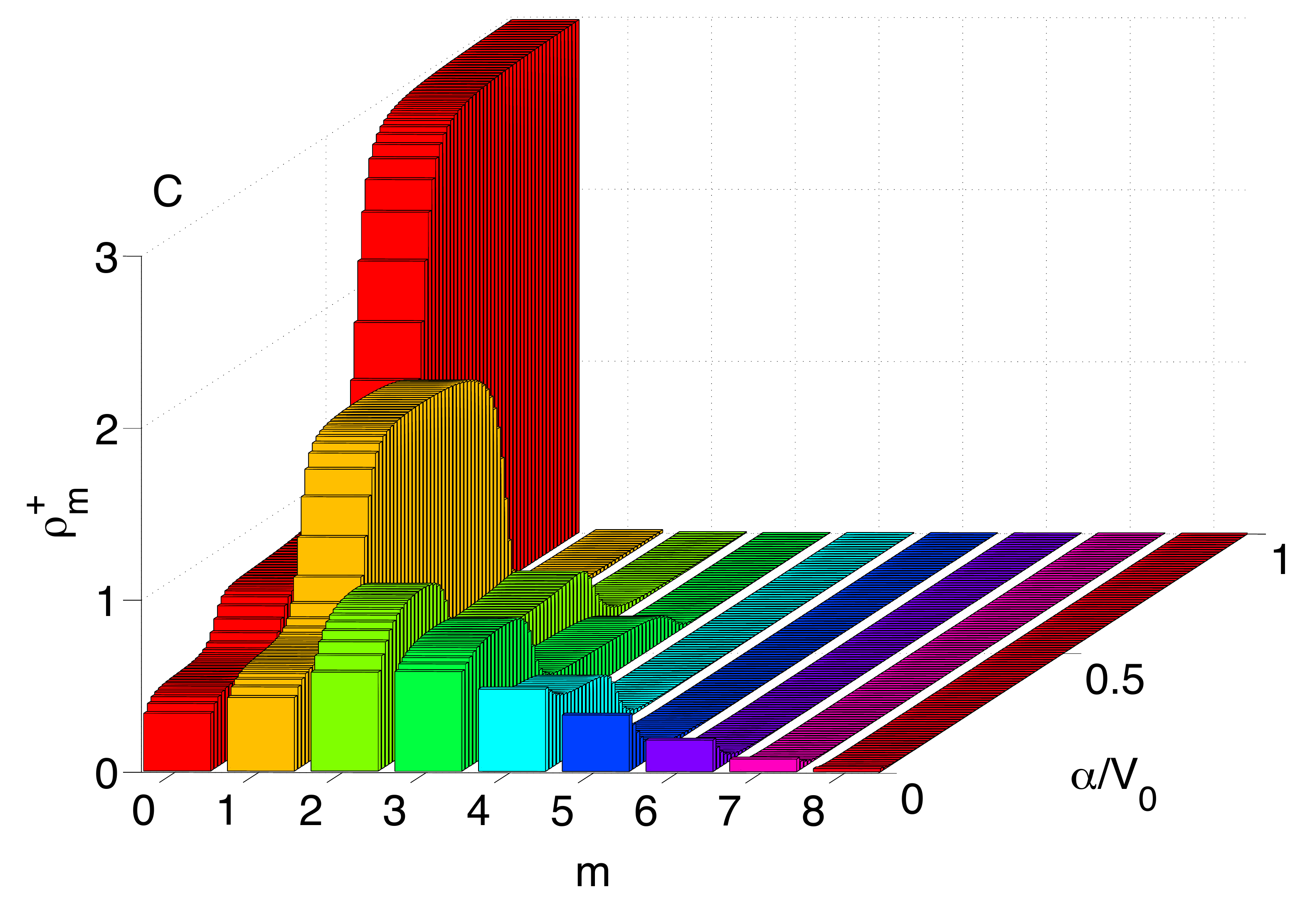}\hspace{0.5cm}
\includegraphics[width=.4\columnwidth]{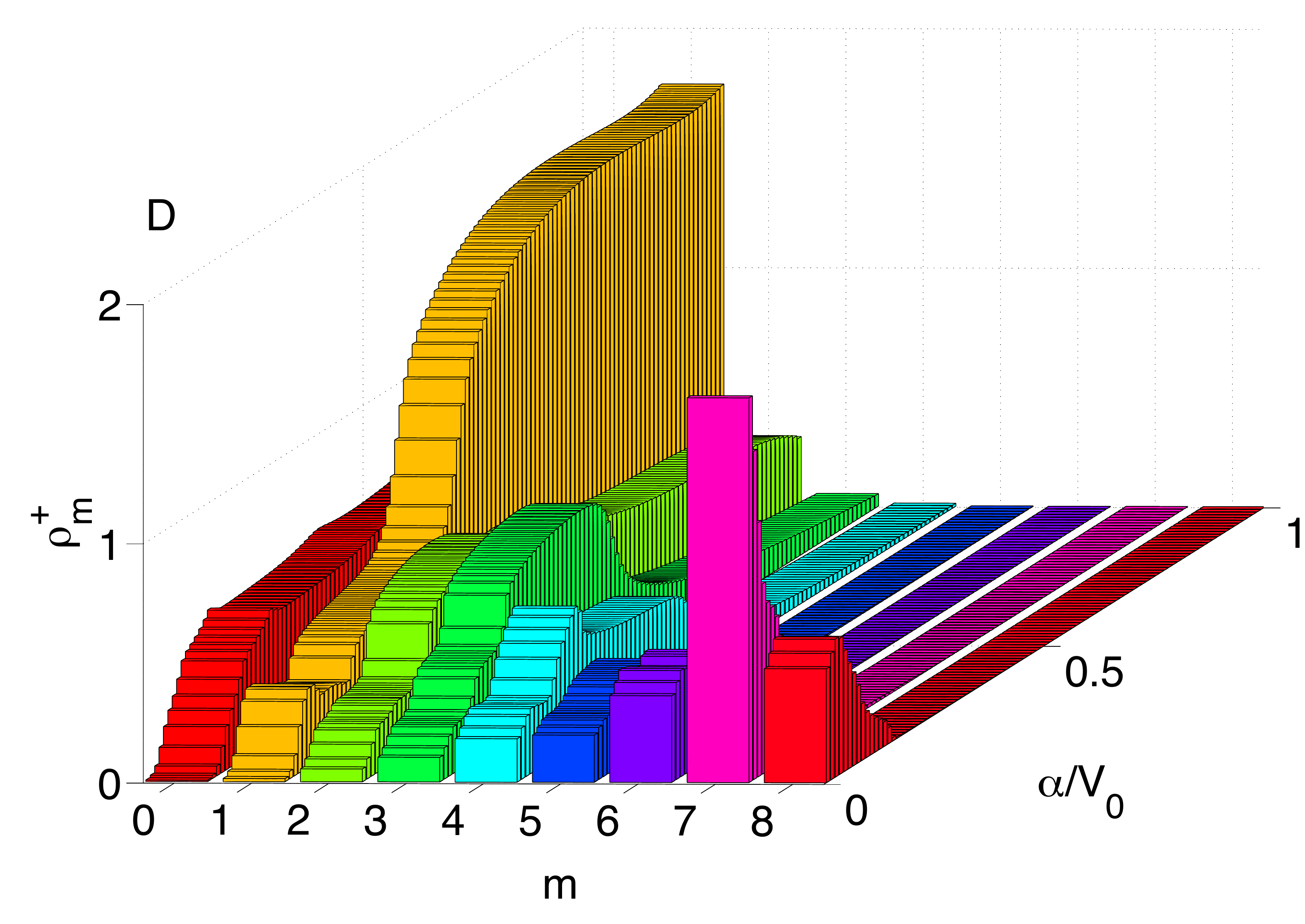}
\end{center}
\caption{\label{fig:momentum3}
One-particle angular-momentum distribution for pseudo-spin $+$ particles
for the ground states of systems whose energy spectra are shown in Fig.~\ref{fig:energy3}.
Compare also with the real-space density profiles shown in Fig.~\ref{fig:densities3}. Note
that the single-particle angular momentum cut-off at $m=10$ defines the sample size for
vanishing $\alpha$ in situations where opposite-spin particles interact (panels B -- D).}
\end{figure}

For our system of interest, an additional possibility arises from the ability to tune the
interaction strength between the two spin components. As seen in panels B and C
of Fig.~\ref{fig:energy3}, a moderate value of $g_{+-}$ turns the crossings occurring
in panel A into anti-crossings, thus, different many-particle states are now
adiabatically connected. Concomitantly, there is a continuous evolution of the
spin-resolved one-particle density profile as a function of the confinement strength
seen in Figs.~\ref{fig:densities3}B and \ref{fig:densities3}C. When interactions
between same-spin and opposite-spin particles have the same magnitude,
the density profile changes significantly (see Fig.~\ref{fig:densities3}D), which
indicates that the character of many-particle ground states is very different from
a fractional-QSH state.

Investigation of the one-particle angular-momentum-state distribution for the few-particle
ground states discussed so far further solidifies our conclusions. See
Figure~\ref{fig:momentum3}. In the absence of interactions between opposite-spin
particles, the characteristic distributions for few-particle versions of the Laughlin and
Laughlin-quasiparticle states emerge at low and intermediate values of $\alpha$.
In the limit of strong trapping potential, the system condenses into the $m=0$ state.
Switching on moderate repulsive (attractive) interaction strength between opposite-spin
particles smoothens the transitions and also shifts the critical values of $\alpha$ to
larger (smaller) values. When interactions among same-spin and between opposite-spin
particles have equal magnitude, the one-particle momentum distribution of the ground
state differs markedly from that associated with a fractional-QH state. It appears that
strong inter-component interactions favor a state with increased occupation of
high-angular-momentum states, spreading out the particles more evenly across
the accessible sample size and leading to an accumulation at the system's boundary.
In the limit of vanishingly small trapping-potential strength $\alpha$, the latter is defined
by the cut-off for single-particle angular momentum applied in our calculations.

\section{Conclusions\label{sec:conc}}

We have considered the interplay of Landau quantization and spin-dependent
interactions in systems where particles with same spin feel the same strong
magnetic field whereas particles with opposite spin are subject to magnetic fields
with the same magnitude but opposite direction. It has been
expected~\cite{ber06a,liu09,lan12} that such systems exhibit the fractional QSH
effect, but we find that interactions between particles with opposite spin weaken
or destroy features associated with fractional-QSH physics. Similar behavior has
been seen in numerical studies of lattice realizations of fractional-QSH
systems~\cite{neu11a}. We have elucidated how behavior that is very different
from ordinary two-component fractional-QH systems is rooted in the drastically
different spectral properties of two-particle interactions for particles feeling the
same versus opposite magnetic-field directions. Thus any feasible route towards
realizing the fractional QSH effect using a spin-dependent uniform magnetic
field~\cite{bee13,ken13} should strive to eliminate interactions between the
opposite-spin components. If the opposite-spin interaction strength is weak, adiabatic
passage between different correlated many-particle states is facilitated by adjusting
the strength of a trapping potential. Our conclusions are supported by numerically
obtained real-space-density profiles and angular-momentum-state occupation
distributions for few-particle systems. The latter could also be utilized as blueprints
for classifying images of correlated ultra-cold atom states.

\section*{Acknowledgments}

Part of the motivation for this project came about from stimulating conversations that one
of us (UZ) had with J.~J.~Heremans and R.~Winkler at the 2011 Gordon Godfrey Workshop
on Spins and Strong Correlations (Sydney, Australia, 24 -- 28 October 2011). We would also
like to thank M.~Fleischhauer and A.~H.~MacDonald for useful discussions. OF was
supported by the Marsden Fund Council from Government funding (contract No.\ MAU1205),
administered by the Royal Society of New Zealand.

\section*{References}


\providecommand{\newblock}{}

\end{document}